\documentclass[fleqn,twoside,twocolumn,nofootinbib,showkeys]{revtex4} 
\usepackage[sec,nocpr]{ujp} 

\begin{document}
\title[Spectrum of Resonance States in $^{6}$He]
{SPECTRUM OF RESONANCE STATES IN \boldmath$^{6}$He. EXPERIMENTAL AND THEORETICAL ANALYSES}
\author{O.M.~Povoroznyk}
\affiliation{Institute for Nuclear Research, Nat. Acad. of Sci. of Ukraine}
\address{47, Prosp. Nauky, Kyiv 03680, Ukraine}
\email{orestpov@kinr.kiev.ua}
\author{V.S.~Vasilevsky\,}%
\affiliation{Bogolyubov Institute for Theoretical Physics, Nat. Acad. of Sci. of Ukraine}%
\address{14b, Metrolohichna Str., Kyiv 03680, Ukraine}%
\email{VSVasilevsky@gmail.com} \udk{539.17} \pacs{21.60.Gx,
24.10.-i, \\[-3pt] 27.20.+n, 25.55.-e} \razd{\secii}

\autorcol{O.M.\hspace*{0.7mm}Povoroznyk,
V.S.\hspace*{0.7mm}Vasilevsky}

\setcounter{page}{201}%

\begin{abstract}
We explore the structure of resonance states in $^{6}$He by
experimental and theoretical methods.\,\,We present the results of
experimental investigations of the three-body continuous spectrum of
$^{6}$He.\,\,For this aim, we use the reaction $^{3}$H$\left(
\alpha,p\alpha\right)nn$, which is induced by the interaction of
alpha-particles with a triton at the beam energy $E_{\alpha} =$~67.2
MeV.\,\,The theoretical analysis of the resonance structure in
$^{6}$He is carried out within the framework of a three-cluster
microscopic model.\,\,The model exploits the hyperspherical
harmonics to describe the intercluster dynamics.\,\,The set of new
resonance states is discovered by the experimental and theoretical
methods.\,\,The energy, width, and dominant decay channels of
resonances are determined.\,\,The obtained results are compared in
detail with the results of different theoretical models and
experiments as well.
\end{abstract}
 \maketitle

\section{Introduction}\vspace*{-0.5mm}

The aim of the present paper is to carry out simultaneous
experimental and theoretical investigations of the resonance
structure of $^{6}$He.\,\,This nucleus has distinguished
three-cluster features.\,\,It has only one loosely bound
state.\,\,The energy of the state is 0.936 MeV with respect to the
three-cluster threshold $\alpha + n + n$.\,\,$^{6}$He belongs to the
so-called Borromean nuclei, which have no bound state in all
two-cluster subsystems.\,\,The ground state has been studied
numerously and thoroughly by different experimental and theoretical
methods.\,\,However, resonance states and continuous spectrum
states, connected with the decay of $^{6}$He onto three clusters,
have not received so much attention.\,\,It is well known that the
investigation of the three-cluster continuous spectrum and resonance
states in this continuum is a challenging problem for experimental
and theoretical physics.\,\,To solve this problem, one needs
sophisticated experimental installations and methods to detect
resonances in the three-cluster continuum and to determine their
quantum numbers (total angular momentum, parity).\,\,From the other
side, one needs advanced theoretical methods to describe correctly
the three-cluster continuum, to reveal resonance states, and to
analyze their properties and nature.

For the correct determination of parameters of unbound excited
states in $^{6}$He, the experimental investigation of the nuclear
reactions should be done, by detecting particles, which testify the
fact of the formation of the excited levels of $^{6}$He, in the
coincidence with one of the decay products (alpha-particle or
neutron) from the resonance structure.\,\,Therefore, we select the
reaction $^{3}$H$(\alpha,p\alpha)2n$, which (as will be shown
latter) is an appropriate tool for detecting the resonance states in
$^{6}$He.\,\,In our experiments, we choose the measurement and
analysis of the $p-\alpha$ coincidence spectra, which result from
the $\alpha+t$ interaction at the beam energy $E_{\alpha}=
67.2$~MeV.\,\,Our choice is based on the successful experimental
study of the four-body reaction $^{3}$H$(\alpha,p\alpha)2n$ \ with
the beam energy $E_{\alpha}=27.2$~MeV \cite{2014MPLA.29.1450105M}.
In \cite{2014MPLA.29.1450105M}, the two-dimensional $E_{p}\times
E_{\alpha }$ spectra obtained in \cite{2001KINR..SP..53G} were
projected on the $E_{p}$-axis in order to give information about the
$^{6}$He resonance states.\,\,In fact, in the case of the $p-\alpha$
coincidence event detection, let $p$ be the nonresonant interacting
particle, and let $\alpha$ be one of three particles constituting
the $^{6}$He three-body resonance.\,\,Then the observation of the
population of such a resonant state can be made by projecting the
yields of the $E_{p}\times E_{\alpha}$ two-dimensional spectrum on
the $E_{p}$-axis.\,\,The energy of the nonresonant  interacting
particle for a given angle is completely determined by that angle
and the resonant energy of the three-particle subsystem.\,\,In this
case, the projection of the $p-\alpha$ coincidence events on the
$E_{p}$-axis gives information about the formed $^{6}$He excited
states, decaying into the $\alpha+n+n$ three-body channel.\,\,From
this experiments \cite{2014MPLA.29.1450105M}, two new low-lying
levels of $^{6}$He at excitation energies of 2.4 and 3 MeV were
detected.\,\,A significant increase in the $\alpha+t$ interaction
energy to $E_{\alpha}= 67.2$~MeV in our new experiment will allow us
to explore the excitation spectrum of $^{6}$He in a more wide energy
range up to 20 MeV, in contrast to the limit equal to 3.5~MeV in the
case of a smaller interaction energy ($E_{\alpha}=27.2$~MeV).

The theoretical analysis of the continuum spectrum in $^{6}$He is
performed within a microscopic three-cluster model, which was
formulated in \cite{2001PhRvC..63c4606V} and used to study resonance
structures in $^{6}$He, $^{6}$Be, and $^{5}$H (see, for instance,
\cite{2010PPN....41..716N}).\,\,Numerous applications of  the  model
demonstrated that it is a reliable and adequate microscopic method
for investigating the three-cluster continuum states  and the
reactions with three clusters in the exit channel  of nuclear
reactions in a wide region of light nuclei.\,\,The model employs
hyperspherical harmonics to numerate channels of the three-cluster
continuum and to implement proper boundary conditions for the decay
of a compound system on three independent clusters.\,\,The large set
of hyperspherical harmonics is involved in calculations in order to
describe different possible scenarios for the decay of a
three-cluster system such as the \textquotedblleft
democratic\textquotedblright\ decay, when all clusters move apart,
or correlated decay, when a selected pair of clusters moves together
with relatively small energy.\,\,The model will be advanced to
include the spin-orbital interaction, which plays an important role
in $^{6}$He, and to calculate different physical quantities, which
help one to understand the nature of three-cluster resonance states.

The key issue of the present paper is the nature of resonance
states, which are embedded in the three-cluster continuum.\,\,Which
is the possible nature of three-cluster resonances, when we have a
very large or infinite set of open channels?\,\,One can expect that
a such resonance state could immediately decay.\,\,However, there
are many resonances states which are experimentally observed in the
three-cluster continuum of light nuclei.\,\,As an example, we can
refer to nuclei with a strong three-cluster feature: $^{6}$He,
$^{6}$Be, $^{9}$B, and so on.\,\,One can suggest two possible
explanations for the appearance of three-cluster
resonances.\,\,First, the resonance state appears in one channel,
which is weakly coupled to other open channels.\,\,The stronger the
coupling of a favorite channel to other ones, the wider is the
resonance width, and, thus, the less are the chances to detect it
experimentally.\,\,The smaller is the coupling of the channels of
the three-cluster continuum, the more narrow resonance state could
be observed.\,\,Second, Baz' \cite{1976JETP...43..205B} demonstrated
that, in many-channel systems, the resonances, which originate from
the distribution of energy over all channels, can appear.\,\,He
called such process as the diffusion-like process.\,\,We are going
to study the nature of resonance states in detail.\,\,We will
determine the dominant decay channels of each resonance state and
the most probable configuration (shape) of a triangle formed by the
centers of mass of three interacting clusters.\,\,We are going to
investigate the resonance states which reside at the energy region
$0\leq E \leq 5$ MeV above the three-cluster $\alpha+n+n$ threshold
or $1 \leq E \leq 6$ MeV of the excitation energy.

The plan of our paper is the following.\,\,In Section
\ref{sect:Exp}, we will explain which reactions and methodology are
used in experimental investigations of the $^{6}$He excited states.
Here, we also demonstrate which results are obtained.\,\,In Section
\ref{sect:Theory}, we will shortly present details of the
microscopic model and make the theoretical analysis of properties of
the $^{6}$He resonance states.\,\,In Section \ref{sect:Compar}, we
compare experimental and theoretical results.\vspace*{-2mm}

\section{Experimental Researches}
\label{sect:Exp}

\begin{figure}%
\vskip1mm
\includegraphics[width=6.0cm]{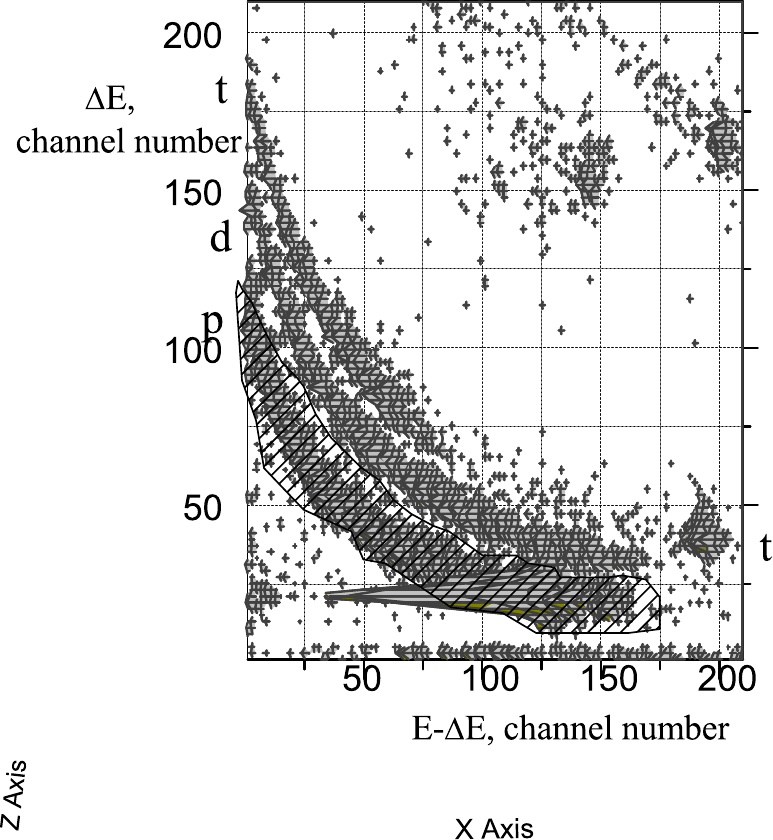}
\vskip-3mm\caption{ Particle distributions in the ($\Delta
E-E$)-plane related to the telescope placed at
$\Theta_{R2}=27.5^{\circ}$ (the right-hand side) with respect to the
beam axis direction.\,\,The locus, composed from the detected
protons, is marked by a dark area }\label{Fig:E02A}
\end{figure}

\begin{figure}%
\vskip1mm
\includegraphics[width=6.0cm]{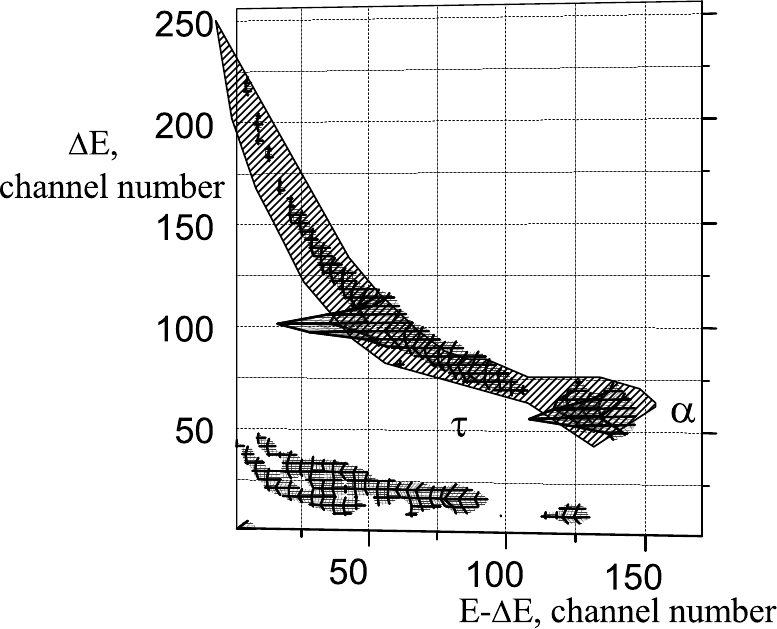}
\vskip-3mm\caption{ The same as in Fig.\,\,\ref{Fig:E02A} but for
the position of the telescope at $\Theta_{L2}= 15^{\circ}$ (the
left-hand side).\,\,The locus, composed from the detected
$\alpha$-particles, is marked by a dark area }\label{Fig:E02B}
\end{figure}

\begin{figure}%
\vskip1mm
\includegraphics[width=6.0cm]{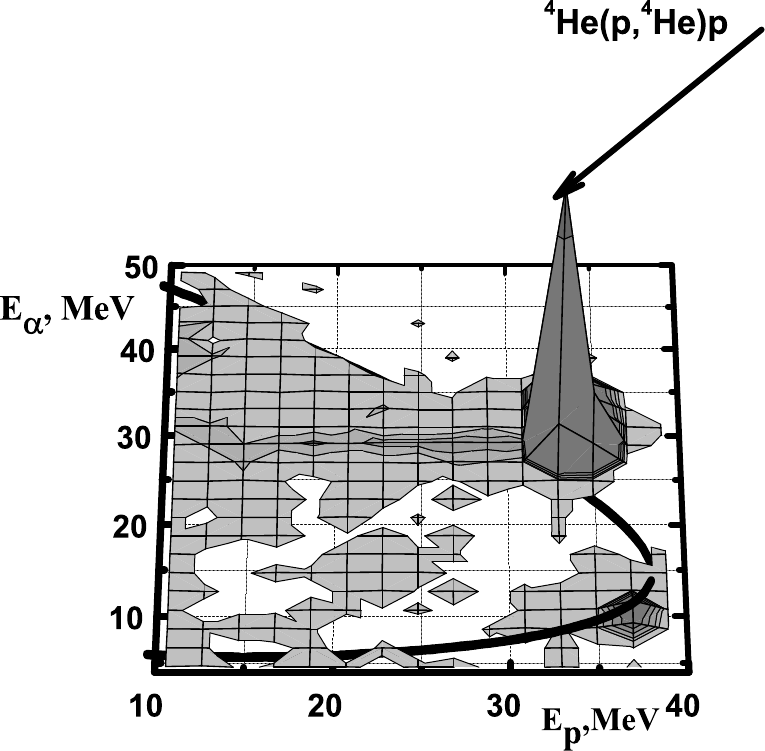}
\vskip-3mm\caption{ Two-dimensional spectra of the
$p\alpha$-coincidences determined in the reactions $^{3}$H$\left(
\alpha,p\alpha\right)  2n$ at $E_{\alpha}=67.2$ MeV.\,\,Arrow points
to events of the elastic $^4$He$+p$ scattering due to hydrogen
impurities in a T-Ti target }\label{Fig:E03}
\end{figure}

The experimental study of the $^{3}$H$(\alpha,p\alpha)2n$ reaction
with four-body exit channel is performed on the isochronous
cyclotron accelerator U-240 of the Institute for Nuclear Research in
Kiev.\,\,The beam energy of $\alpha$-particles is determined to be
$E_{\alpha} = 67.2 \,\pm$ $\pm\, 0.4$~MeV by using a technique
developed to measure the time and energy characteristics of the
cyclotron beam \cite{1991KINR..Preprint..91Z}.\,\,For this aim, we
also use a pair of $\Delta E-E$ telescopes to detect the proton and
$\alpha$-particle coincidence from the four-body
$^{3}$H$(\alpha,p\alpha)2n$ reaction.\,\,The first telescope was
placed at the right-hand side and consisted of $\Delta E$ [400
$\mu$m thick totally depleted silicon surface barrier detector
(SSD)] and $E$ [NaI(Tl) with 20 mm $\times$ 20 mm t]
detectors.\,\,The second telescope was placed at the left-hand side
and consisted of $\Delta E$ (90 $\mu$m SSD) and $E$ [Si(Li) with
3~mm] detectors.\,\,The first telescope can detect protons,
deuterons, and tritons (see Fig.~\ref{Fig:E02A}), whereas the second
telescope can detect $\tau$-and $\alpha$-particles (see
Fig.~\ref{Fig:E02B}) together with protons, deuterons, and tritons
of low energies.\,\,The calibration of scintillators is made by
using the procedure described in our previous paper
\cite{2011JPSJ...80i4204P}, while a standard technique is used for
the SSD.\,\,We record the signals coming from two telescopes within
a window time of about 100 ns, by using a standard electronic
set-up.\,\,One of the $E_{p}\times E_{\alpha}$ two-dimensional
spectra of the $p-\alpha$ coincidence events is presented in Fig.
\ref{Fig:E03}.

\begin{figure}%
\vskip1mm
\includegraphics[width=7.5cm]{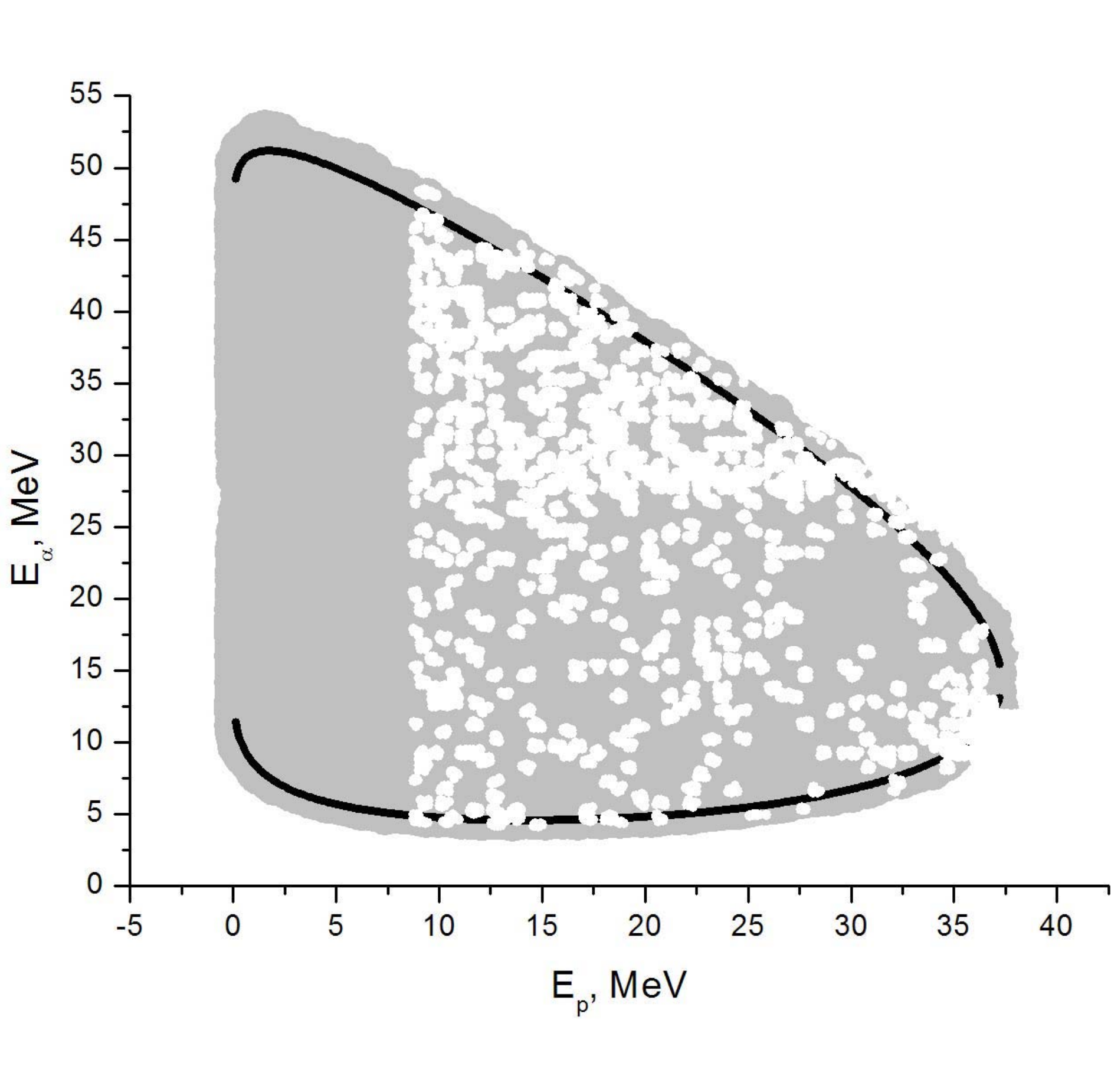}
\vskip-3mm\caption{ Selected experimental two-dimensional
spectrum of the
$p\alpha$-coincidence (white dots) 
 and corresponding kinematical calculations }\label{Fig:E04}
\end{figure}

\begin{figure}%
\vskip3mm
\includegraphics[width=7cm]{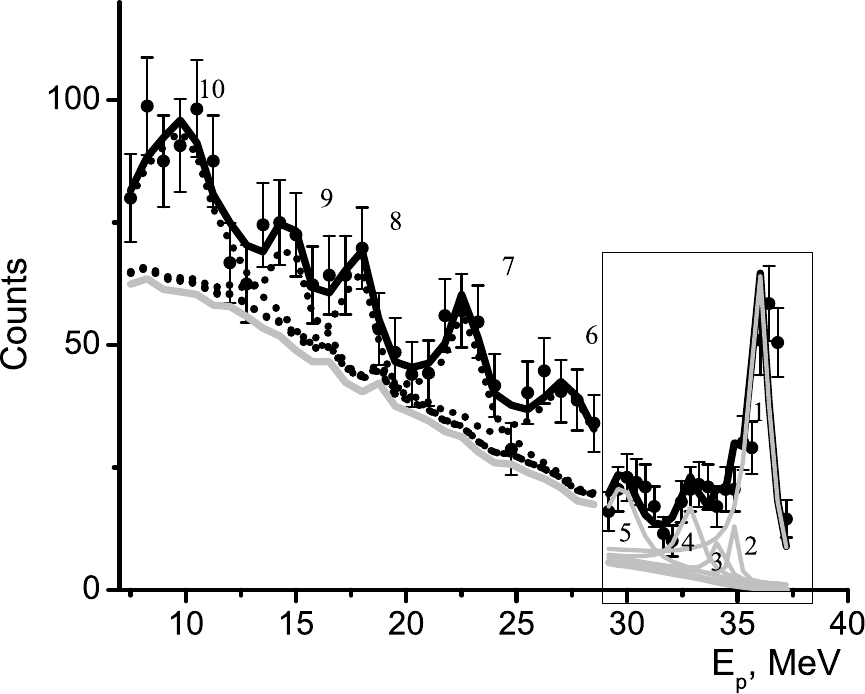}
\vskip-3mm\caption{ Resonance structure of $^{6}$He discovered  in
the reaction $^{3}$H$(\alpha,p\alpha)2n$.\,\,The proton energy
$E_{p}$ determined from the two-dimensional spectra of
$p\alpha$-coincidences is projected on the excitation energy of
$^{6}$He.\,\,The results of approximation represented by expression
(\ref{eq:E001}) are plotted by black solid lines.\,\,The
contributions of different excited levels of $^{6}$He are marked by
numbers {\it 1--10} and displayed by dashed lines.\,\,The low-energy
part of experimental data is shown in a frame }\label{Fig:E05A}
\end{figure}

\begin{figure}%
\vskip3mm
\includegraphics[width=7cm]{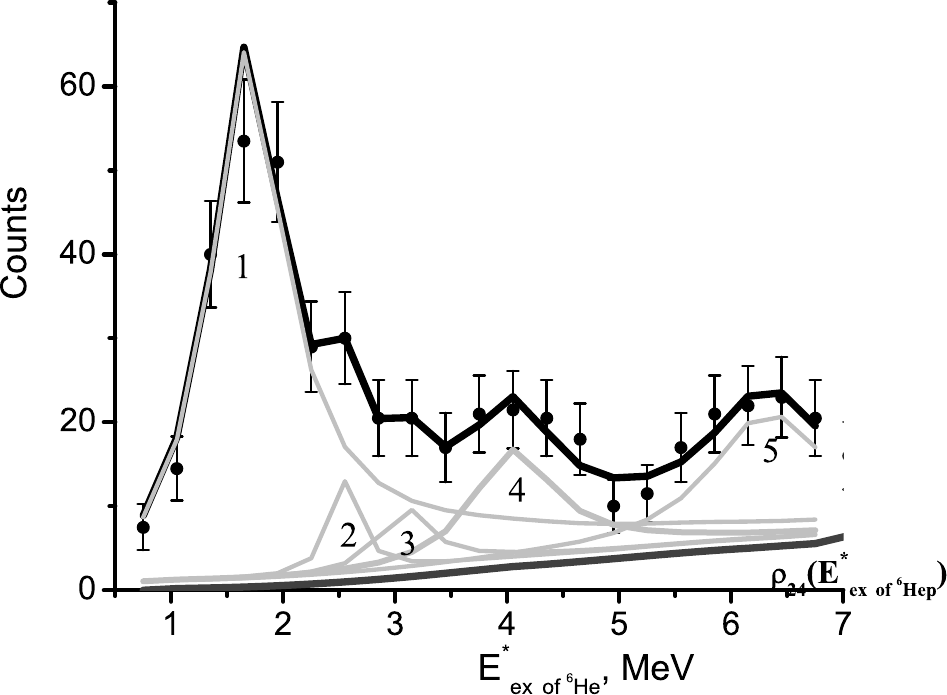}
\vskip-3mm\caption{ The same as in Fig.\,\,\ref{Fig:E05A} but for
the low-energy spectrum of $^{6}$He excitations }\label{Fig:E05B}
\end{figure}

The further analysis consists in the selection of coincidence
$p\alpha$-events from the four-body $^{3}$H$(\alpha,p\alpha)2n$
reaction in the measured two-dimensional $E_{p}\times E_{\alpha}$
spectrum.\,\,It is well known that the $p\alpha$-coincidence events
emerging from the four-body $^{3}$H$(\alpha,p\alpha)2n$ are disposed
in parts of the represented spectrum, which are bounded by kinematic
loci (see Fig.\,\ref{Fig:E03}) and calculated for the hypothetical
three-body $^{3}$H$(\alpha,p\alpha)\left\langle 2n\right\rangle $
reaction, where $\left\langle 2n\right\rangle $ is a particle
consisting of two neutrons, which move in one direction with the
energy of relative motion equal to zero \cite{1972NucIM..98..301F}.
The further analysis consists in the correct selection of these
$p\alpha$-coincidence events.\,\,First, by using the Monte
Carlo simulation procedure 
\cite{2014MPLA.29.1450105M, 2007NPAE...131P, 2012PhRvC..85f4330P}
and by accounting for the real experimental conditions (the value of
energy beam blurring, the dimensions of solid angles of detectors
and beam spot), we determine a part of the ($E_{p}\times
E_{\alpha}$) plane (Fig.\,\ref{Fig:E04}, light grey background), in
which the four-body experimental events can manifest
themselves.\,\,For each $i$-th light grey dot with $E_{\rm
pkini}\times E_{\alpha {\rm kini}}$ coordinates with regard for the
energy and momentum conservation laws \cite{2000PhRvL..85..262N,
1984PhRvC..30..158R}, we calculate the value $Q_{2-4{\rm
kini}}$.\,\,We use the Monte Carlo method for drawing the
experimental $E_{p}\times E_{\alpha}$ spectrum (Fig.\,\ref{Fig:E03})
and, for every obtained $E_{\rm piexp}\times E_{\alpha {\rm iexp}}$
event (see
Fig.\,\ref{Fig:E04}, white dots), determine the values $Q_{2-4{\rm expi}}$.\,\,
The criterion for selecting the necessary experimental events, which
was formulated in \cite{2014MPLA.29.1450105M}, is used to extract
them from the total array of experimental points.\,\,The white dots
in Fig.\,\ref{Fig:E04} are the selected experimental coincidence
$p\alpha$-events obtained from the analysis of the four-body
$^{3}$H$(\alpha,p\alpha)2n$ reaction.\,\,More details about the
implementation of the Monte Carlo procedure and the definitions of
all input and output variables can be found in Ref.
\cite{2007NPAE...131P}.\,\,The white dots presented in
Fig.\,\ref{Fig:E04} are the selected experimental coincidence
$p\alpha$-events obtained as a result of the study of the four-body
$^{3}$H$(\alpha,p\alpha)2n$ reaction.\,\,In this figure, it is
evident that the experimental events are cut at $E_{p}< 10$ MeV,
which is due to the use of $\Delta E$-detectors about 400 $\mu$m in
thickness in the left telescope to detect and to identify protons.
For the four-body reaction, resonances in two-body subsystems
produce an increase in the intensity of break-up events in those
places of the $E_{p}\times E_{\alpha}$ spectra, where the
corresponding energy of the relative motion of decayed clusters
achieves resonance energies.\,\,We observed no resonance phenomenon
caused by the $\alpha+p$ interaction in the obtained experimental
data.\,\,But before the analysis of two-dimensional $E_{p}\times
E_{\alpha}$ spectra from the $\alpha+t$ interaction at $E_{\alpha} =
27.2$ MeV \cite{2012PhLB..718..441M}, one can observed some bars
filled by events, which are parallel to the alpha-particle energy
axis, and these strips have been identified as a manifestation of
the formation of the excited states in nucleus $^{6}$He and their
subsequent decay by emitting three clusters $\alpha+n+n$.

Since we can explore the much higher excitation energy ($E_{\alpha}
= 67.2$ MeV) in this experiment, we hope for that this study will
repeat the previous experiment and enhance our knowledge of the
structure of the high-energy part of the excitation spectrum of
$^{6}$He.\,\,Indeed, in the spectrum (see Fig.\,\ref{Fig:E05A})
resulting from the projection of $p\alpha$-coincidence events of the
four-body $^{3}$H$(\alpha ,p\alpha)2n$ reaction (all white dots in
the limits of the light grey background) on the proton energy axis,
we can observe numerous resonant manifestations.\,\,By our
assumption, there are two possible mechanisms of this nuclear
transformation~-- instantaneous formation of these four particles
(so-called statistical model) and two-stage process of formation
$p$, $\alpha $-particle, and two neutrons, in which a proton and
$^{6}$He nucleus in one of the excited states are formed on the
first stage.\,\,On the second stage, the nucleus decays on three
parts~-- $\alpha$-particle and two neutrons.\,\,Then the yield of
the four-body reaction can be calculated by a sum of the
Breit--Wigner contributions with regard for the contribution of the
simultaneous formation of proton, alpha-particle and two neutrons in
the output channel of this reaction:
\begin{equation}\label{eq:E001}
N\propto\rho_{24}\left(  E_{p}\right) \! \sum_{j=1}^{k}\left[
C_{j}\frac {\Gamma_{j}/2}{\left(  E_{j}-E_{p}\right)  ^{2}+\left(
\Gamma_{j}/2\right) ^{2}}+B\right]\!,\!\!
\end{equation}
where $\rho_{24}\left(  E_{p}\right)$ is the projection of the phase
space factor on the $E_{p}$ axis for the detection on coincidence of
protons and alpha-particles from the four-body
$^{3}$H$(\alpha,p\alpha)2n$ reaction; $C_{j}$ is the corresponding
contribution of each unbound $^{6}$He$^*$ state decaying into the
$\alpha+n+n$ channel, and $B$ is the corresponding contribution due
to the simultaneous formation of $p$, $\alpha,$ and two $n$.\,\,The
values of excitation energy of $^{6}$He and the phase space factor
of the sequential three-body reaction, used in expression
(\ref{eq:E001}), are calculated with regard for the geometry and the
energy parameters of the experiment and by using the Monte Carlo
simulation.\,\,The fitting procedure is carried out with the
least-square method, and the quantities $E_{j}$, $\Gamma_{j}$,
$C_{j},$ and $B$ were as variables.\,\,The obtained energy
parameters of ten excited levels from the fitting procedure are
reported \mbox{in Table~\ref{Tab:Table1}.}

Thus, we discovered ten excited levels of $^{6}$He (see also Fig.
\ref{Fig:E05A}).\,\,For each level, we determined the excitation
energy and the width (see Table \ref{Tab:Table1}).\,\,Most part of
the discovered resonance states are narrow ones.\,\,Their total
width varies from $0.4\pm 0.2$ to $2.3\,\pm$ $\pm\, 1.0$ MeV and is
much less than the resonance energy measured from the three-cluster
$\alpha+n+n$ threshold.

\begin{table}[b]
\noindent\caption{ Excitation energy \boldmath($E$) and the width
($\Gamma$)\\ for the excited levels obtained in reactions generated\\
by the $^3$H${}+\alpha$ interaction.\,\,Index $N_s$ numerates\\
excited levels in $^6$He.\,\,$R1$ represents the reaction\\
$^3$H${}+ \alpha\rightarrow p+{}^6$He${} \rightarrow p+\alpha+n+n$
and $R2$ stands\\ for the reaction $^3$H${}+\alpha \rightarrow
p+{}^6$He${} \rightarrow p+t+t$ }\vskip3mm\tabcolsep1.3pt

\noindent{\footnotesize\begin{tabular}{|c|c|c|c|c|c|} \hline
\multicolumn{3}{|c}{\rule{0pt}{5mm}$R1$, $E_{\alpha}=67.2$ MeV} &
\multicolumn{3}{|c|}{$R1$,  $E_{\alpha}=27.2$ MeV,
\cite{2014MPLA.29.1450105M}}\\[2mm]
\hline
\multicolumn{3}{|c}{\rule{0pt}{5mm}$\Theta_{p}/\Theta_{\alpha}=27.5^{\circ}/$15.0$^{\circ}%
$} & \multicolumn{3}{|c|}{$\Theta_{p}/\Theta_{\alpha}=28.5^{\circ}/13.0^{\circ}\!\!+\!16.5^{\circ}\!\!+\!19.5^{\circ}$}\\[2mm]
  \hline \rule{0pt}{5mm}$N_{s}$ & $E$, MeV & $\Gamma$, MeV & $N_{s}$ & $E$, MeV & $\Gamma$, MeV
  \\[2mm]
   \hline
  \rule{0pt}{5mm}~\,1 & ~\,$1.7\pm 0.2$ & $0.65\pm 0.15$ & ~\,1 & $1.8\pm 0.2$ & ~\,$0.3 \pm 0.15$ \\
  ~\,2 & ~\,$2.5 \pm 0.2$ & $0.4\pm 0.2$ & ~\,2 & $2.4 \pm 0.2$ & $0.4\pm 0.2$ \\
  ~\,3 & ~\,$3.1 \pm 0.3$ & $0.4 \pm 0.2$ & ~\,3 & $3.0 \pm 0.2$ & $0.6 \pm 0.2$ \\
  ~\,4 & ~\,$4.1 \pm 0.3$ & $0.9 \pm 0.3$ & \multicolumn{3}{|c|}{$R2$, $E_{\alpha}=67.2$ MeV,
  \cite{2012PhRvC..85f4330P}} \\
  ~\,5 & ~\,$6.1 \pm 0.3$ & $1.6 \pm 0.3$ & 10 & $18.3 \pm 0.2$ & $0.0 \pm 0.2$ \\
  ~\,6 & ~\,$8.8 \pm 0.4$ & $2.0 \pm 0.6$ &
\multicolumn{3}{|c|}{$R2$, $E_{\alpha}=67.2$ MeV, \cite{2012PhRvC..85f4330P}} \\
  ~\,7 & $11.6 \pm 0.4$ & $2.0 \pm 0.7$ & \multicolumn{3}{|c|}{$\Theta_{p}%
/\Theta_{t}=21.0^{\circ}/15.0^{\circ}$, 21.0$^{\circ}/$20.0$^{\circ}
$} \\
  ~\,8 & $14.6 \pm 0.4$ & $2.3 \pm 1.0$ & ~\,8 & $14.0 \pm 0.4$ & $0.6 \pm 0.4$ \\
  ~\,9 & $16.4 \pm 0.4$ & $1.4 \pm 0.9$ & ~\,9 & $16.1 \pm 0.4$ & $0.8 \pm 0.4$ \\
  10 & $18.5 \pm 0.4$ & $1.7 \pm 0.6$ & 10 & $18.4 \pm 0.4$ & $1.0 \pm 0.4$ \\[2mm]
  \hline
\end{tabular}\label{Tab:Table1}}
\end{table}

Note that, in our experiments, we discovered quite new resonance
states in $^6$He and confirmed the existence of some resonance
states, which have been observed in other experiments.\,\,For
instance, the energy parameters of the observed excited states,
marked as $N_{s}=2$ and $N_{s}=3$, almost coincide within
experimental errors with the results of research at the cyclotron
U-120 \cite{2014MPLA.29.1450105M}.\,\,The excited level broader than
our $N_{s}=2$ state was observed recently in the experiment using a
beam of radioactive nuclei $^{8}$He
\cite{2012PhLB..718..441M}.\,\,The excited states with energy
parameters close to those of our levels $N_{s}=8$, $N_{s}=9,$ and
$N_{s}=10$ were observed in the inclusive deuteron spectra
determined from the $^{7}$Li$(n,d)^{6}$He reaction
\cite{1977PhRvC..16...31B} and from measurements of the $t-t$ and
$p-t$ coincidence events in the $^{3}$H$(\alpha,tt)p$ and
$^{3}$H$(\alpha,pt)t$ reactions \cite{2012PhRvC..85f4330P}.\,\,The
energy parameters of our $N_{s}=4$ excited state are close to the
experimental values from Ref.\,\,\cite{2001PhLB..518...27L}, where
the excitation spectrum of $^{6}$He has been studied in the
$^{6}$He$(p,p\alpha)$ reaction.\,\,The resonances almost similar to
our states $N_{s}=3$, 5, and 6 were observed in the experimental
study of the $^{7}$Li$(t,\alpha)^{6}$He reaction
\cite{1954PhRv...96..684A}, which is one of the first experiments
devoted to the study of the excitation spectrum of $^{6}$He.
Recently, Mougeot {\it et al.} \cite{2012PhLB..718..441M} \ observed
two new resonance states.\,\,The resonances with the parameters
$E=2.6\pm 0.3$~MeV,  $\Gamma=1.6\pm 0.4$~MeV, $J^{\pi}=2^{+}$ and
$E=5.3\pm 0.3$~MeV,  $\Gamma=2\pm 1$~MeV, $J=1$ were discovered in
the transfer reaction $p\left(^{8}{\rm He},t\right)$ induced by a
beam of radioactive nuclei $^{8}$He.\,\,As is seen, the energy of
the first resonance state is very close to that of our $N_{s}=2$
state.\,\,However, its width is much larger than that deduced in our
experiments.\,\,The second resonance state, discovered in
\cite{2012PhLB..718..441M}, lies between our $N_{s}=4$ and $N_{s}=5$
states, not far from the $N_{s}=5$ state.

The comparison of different experimental methods demonstrates
the consistency of our experimental results with results of other
experimental investigations of the $^{6}$He resonance structure.

\section{Theoretical Research}
\label{sect:Theory}

In this section, we briefly present the main ideas of a microscopic
model formulated in Refs.
\cite{2001PhRvC..63c4606V,2001PhRvC..63c4607V}.\,\,Note that a
similar model was suggested in \cite{2004NuPhA.740..249K}, which
employs the generator coordinate method and was recently
\cite{2009PhRvC..80d4310D} used to study the resonances in $^{6}$He.

The present model is an extension of the resonating group method for
the description of three-cluster systems.\,\,It combines the
$J$-matrix method or the algebraic version of the resonating group
method with the hyperspherical harmonics (HH) method.\,\,The former
indicates that we use the full set of square integrable functions
(namely, a specific set of three-cluster oscillator functions) to
reduce the Schr\"{o}dinger equation to that in the matrix form and
to solve it.\,\,It is well known that the hyperspherical harmonics
method \cite{2001PhRvC..63c4606V, 2001PhRvC..63c4607V,
 1993PhR...231..151Z, 1996PhR...264...27B,
 1998NuPhA.631..793C, 1998PhRvC..58.1403C,
 2010PPN....41..716N, 2004AnPhy.312..284L} provides simple tools
for imposing the suitable boundary conditions for the discrete and continuous
spectra of three-cluster systems.

A microscopic model is based on two key elements: a Hamiltonian,
which consists of the kinetic energy operator and the sum of
pairwise nucleon-nucleon interactions, and a wave function.\,\,The
later determines, which part of the total Hilbert space is involved
in the description of a many-particle system. That is why we start
with a wave function of~$^{6}$He.

In the present three-cluster model, the wave function of $^{6}$He can be
written as
\[\Psi_{JM} =\]\vspace*{-7mm}
\begin{equation}\label{eq:323_07}
=\sum_{SL}\widehat{\mathcal{A}}\left\{  \left[ \Phi_{1}\!\left(
\alpha,b\right)  \Phi_{2}\!\left(  n\right) \Phi_{2}\!\left(
n\right) \right] _{S}\phi_{L}\!\left(
\mathbf{q}_{1},\mathbf{q}_{2}\right) \right\}  _{JM}\!,\!\!\!
\end{equation}
where $\Phi_{1}\left(  \alpha,b\right)  $ is an antisymmetric
shell-model wave function describing the internal structure of the
alpha-particle with four nucleons in the $s$-shell.\,\,We indicate
explicitly that the wave function of the alpha-particle
$\Phi_{1}\left( \alpha,b\right)$, being an eigenfunction of the
many-particle Hamiltonian with the harmonic oscillator interaction,
depends on the oscillator length $b$, which is a variational
parameter and has to be fixed (selected) in numerical
calculations.\,\,The neutron wave function $\Phi _{2}\left(n\right)$
only includes the spin and isospin variables of the neutron.\,\,The
quantity $\widehat{\mathcal{A}}$ stands for the total
antisymmetrization operator, and the vectors $\mathbf{q}_{1}$ and
$\mathbf{q}_{2}$ are the Jacobi vectors determining a relative
position of clusters in the coordinate space.

The intercluster wave function $\phi_{L}\left(  \mathbf{q}_{1},\mathbf{q}%
_{2}\right)  $ of relative three-cluster motion is to be determined
by solving the Schr\"{o}dinger equation with the proper boundary
conditions.\,\,For this aim, we introduce the hyperspherical
coordinates $q_{1}=\rho\sin\theta$ and $q_{2}=\rho\cos\theta$ and
expand the wave function $\phi_{L}\left(\mathbf{q}_{1},\mathbf{q}
_{2}\right)$ in the hyperspherical harmonic basis
\[\phi_{L}\left(  \mathbf{q}_{1},\mathbf{q}_{2}\right)
=\!\sum_{l_{1},l_{2} }\phi_{l_{1},l_{2};L}\!\left(
\rho,\theta\right)
\left\{  Y_{l_{1}}\!\left(\widehat{\mathbf{q}}_{1}\right)  Y_{l_{2}}\!\left(  \widehat{\mathbf{q}}%
_{2}\right)  \right\}  _{LM}=\]\vspace*{-7mm}
\begin{equation}\label{eq:323_08}
=\sum_{K, l_{1}, l_{2}}\!\phi_{K;l_{1},l_{2};L}\!\left(  \rho\right)
\chi _{K}^{\left(  l_{1},l_{2}\right)  }\!\left(  \theta\right)
\left\{ Y_{l_{1} }\!\left(  \widehat{\mathbf{q}}_{1}\right)
Y_{l_{2}}\!\left( \widehat{\mathbf{q}}_{2}\right)  \right\}  _{LM},
\end{equation}
where $\widehat{\mathbf{q}}_{1}$ and $\widehat{\mathbf{q}}_{2}$\ are
the unit vectors.\,\,The explicit form of the hyperspherical
harmonics $\chi_{K}^{\left(  l_{1} ,l_{2}\right)  }\left(
\theta\right)  $ and the set of equations for hyperradial
excitations $\phi_{K;l_{1},l_{2};L}\left(  \rho\right)$ can be found
in \cite{2001PhRvC..63c4606V, 2001PhRvC..63c4607V,
2010PPN....41..716N}.

The hypermomentum $K$ and the partial angular momenta $l_{1}$ (along
$\mathbf{q}_{1}$) and $l_{2}$ (along $\mathbf{q}_{2}$) define the
three-cluster geometry and characterize the different scattering
channels.\,\,These three quantum numbers will be denoted as
$c=\left\{ K;l_{1} ,l_{2}\right\}$.\,\,\ In the case where the total
spin $S$ and the total orbital momentum $L$ of the compound system
are not good quantum numbers (this takes place when, e.g., the
spin-orbital components are presented in the nucleon-nucleon
potential and are involved in calculations), five quantum numbers
will numerate channels of the three-cluster system $c=\left\{
K;l_{1},l_{2};LS\right\} $.\,\,Within the present model, the total
spin $S$ of $^{6}$He coincides with the spin of the two-neutron
subsystem and may have two values $S=0$ and $S=1$.\,\,The value
$S=0$ is dominant, because it realizes a more stronger interaction
between valence neutrons.\,\,We recall that two neutrons with $S=0$
and the zero orbital momentum have a virtual state, which strongly
enhances the cross section of elastic neutron-neutron scattering.

The hyperradial functions $\phi_{K;l_{1},l_{2};L}\left(\rho\right)$
at large values of $\rho$ have the asymptotic form
\[\phi_{K;l_{1},l_{2};L}\left(  \rho\right)=\phi_{c;L}\left(  \rho\right)\simeq\]\vspace*{-7mm}
\begin{equation}\label{eq:021}
\simeq \delta_{c_{0},c}\psi_{c}^{\left(  -\right)}\left(
k\rho\right) -S_{c_{0},c}\psi_{c}^{\left(  +\right)  }\left(
k\rho\right),
\end{equation}
where $c_{0}$ denotes the incoming channel, $S_{c_{0},c}$  is an
element of the many-channel $S$-matrix, and $\psi_{c}^{\left(
+\right) }\!\left( k\rho\right)$ ($\psi_{c}^{\left(-\right)}\!\left(
k\rho\right) $) is the incoming (outgoing) wave.\,\,(The definitions
of the wave functions $\psi _{c}^{\left(+\right)}\left( k\rho\right)
$ and $\psi_{c}^{\left( -\right)}\left( k\rho\right)$ can be found
in \cite{2001PhRvC..63c4606V}).\,\,Equation (\ref{eq:021}) shows
explicitly the boundary conditions for the wave function of
three-cluster continuous spectrum states.\,\,These boundary
conditions are implemented in the system of equations for the
channel functions $\phi_{K;l_{1},l_{2};L}\left(\rho\right)$.\,\,By
solving these system, we obtain the wave functions of the
three-cluster continuum and the $S$-matrix, which describe all kinds
of elastic and inelastic processes in the compound system.

For the numerical calculation of properties of $^{6}$He, we make use
of two effective NN-potentials: the Volkov (VP) \cite{kn:Volk65} and
Minnesota potentials (MP) \cite{kn:Minn_pot1}.\,\,These two
potentials are frequently used in various microscopic models for
investigating the light atomic nuclei ($^{6}$He, in
particular).\,\,The Volkov potential contains only central
components of the $NN$ interaction, and, thus, the total spin $S$
and the total orbital momentum $L$ are quantum numbers, as both $S$
and $L$ are the integrals of motion.\,\,For this potential, we
consider only one value of total spin $S=0$.\,\,Since the Minnesota
potential contains spin-orbital components (we took the IV-th
version of the spin-orbital forces from Ref.
\cite{1970NuPhA.158..529R}), the total spin $S$ \ and the total
orbital momentum $L$ are no longer good quantum numbers.\,\,Thus,
the total angular momentum $J$ in the general case can be presented
by a combination of four states \ ($L,S$) with different values of
the total spin $S$ and the total angular momentum $L$:
\[
\left\vert J\right\rangle =\left\vert \left(  J,0\right)
\right\rangle +\left\vert \left(  J-1,1\right)  \right\rangle
+\left\vert \left( J,1\right)  \right\rangle +\left\vert \left(
J+1,1\right)  \right\rangle\!.
\]
For $J^{\pi}=0^{+}$ and $J^{\pi}=1^{-}$ angular momenta, we have the
following combination: $\left\vert 0^{+}\right\rangle  = \left\vert
\left(  0,0\right) \right\rangle +\left\vert \left(  1,1\right)
\right\rangle$ and $\left\vert 1^{-}\right\rangle = \left\vert
\left(  1,0\right) \right\rangle +\left\vert \left(  1,1\right)
\right\rangle +\left\vert \left(  2,1\right)  \right\rangle $.

In the present model, we have got several input parameters such as
the oscillator radius (or length) $b$ and the maximal value of
hypermomentum $K_{\max}$ (which determines the maximal number of
channels of the three-cluster continuum).\,\,We select the
oscillator length to minimize the alpha-particle energy.\,\,For the
Volkov and Minnesota potentials, this can be achieved with
$b=1.37$~fm and $b=1.285$ fm, respectively.\,\,For positive parity
states, we take $K_{\max}=14.$ For negative parity states, we
involve all hyperspherical harmonics up to $K_{\max}=13$.\,\,Such
set of hyperspherical harmonics provides stable results for the
energy and the width of resonance states (see details in
\cite{2001PhRvC..63c4607V, 2001PhRvC..63f4604V}).\,\,It was
demonstrated in \cite{2001PhRvC..63f4604V} that these hyperspherical
harmonics account for the numerous scenarios of the decay of the
three-cluster system.\,\,We will use a large space of hyperradial
excitations $n_{\rho.\max}=70$, which is equivalent to
140$\hbar\Omega$ excitations in the shell model.\,\,This value of
$n_{\rho.\max}$ is sufficiently large to provide the correct
description of the internal part of the three-cluster wave function
and to reach the asymptotic region.

We start our investigation with the calculation of the ground-state
energy.\,\,In Table~2, we display the ground-state energy of
$^{6}$He and the parameters $m$ and $u$ for the Volkov and Minnesota
potentials, respectively.\,\,In Table~2, we also indicated the
values of oscillator length~$b$.

\begin{table}[b]
\noindent\caption{ Input parameters, adjustable\\ parameters of the
potential, and bound\\ state energy of \boldmath$^6$He
}\vskip3mm\tabcolsep2.3pt

\noindent{\footnotesize\begin{tabular}{|l|c|c|c|c|c|}
  \hline
\multicolumn{1}{|c}{\rule{0pt}{5mm}Potential} &
\multicolumn{1}{|c}{VP N1 } & \multicolumn{1}{|c}{VP N2 } &
\multicolumn{1}{|c}{VP N2 } & \multicolumn{1}{|c}{MP } &
\multicolumn{1}{|c|}{Exp } \\[2mm]
  \hline
  \rule{0pt}{5mm}$b$, fm & 1.37 & 1.37 & 1.37 & 1.285 & $-$\\
  Parameter & $m=0.60$ & $m=0.60$ & $m=0.615$ & $u=0.98$ & $-$ \\
  $E$, MeV & $-1.387$ & $-1.234$ & $-0.953$ & $-0.551$ & $-0.973$ \\[2mm]
  \hline
\end{tabular}}
\end{table}

One can see that the Volkov potentials N1 and N2 with the original
value of Majorana parameter $m=0.60$ overbound the ground state of
$^{6}$He; whereas the Minnesota potential with the parameter
$u=0.98$, which provides a correct description of the $\alpha+n$
subsystem, underbound the ground state of $^{6}$He.\,\,The same
values of parameter $u$ and oscillator radius $b$ were used in Refs.
\cite{1993PhRvC..48..165C} to study the bound states of $^{6}$He,
$^{6}$Li, and resonance states in $^{6}$He, $^{6}$Li, and $^{6}$Be
within the complex scaling method.\,\,For the Volkov potential N2,
we found the value of Majorana parameter ($m=0.615$), which gives
the proper ground-state energy.

By solving the system of dynamic equations, we obtain the scattering
S-matrix $\left\Vert S_{cc^{\prime}}\right\Vert $ for the $N_{\rm
ch}$ channel system and the $N_{\rm ch}$ wave functions.\,\,The
element $S_{c_{,}c^{\prime}}$ of the $S$-matrix describes the
transition from the initial channel $c$ to the final channel
$c^{\prime}$.\,\,To analyze the states of the three-cluster
continuum and transitions in the continuum, we use two different
representations of the $S$-matrix.\,\,In the first representation,
we determine the inelastic parameter $\eta_{c_{,}c^{\prime}
}$ and the corresponding phase shift $\delta_{c_{,}c^{\prime}}$%
\begin{equation}\label{eq:005}
S_{c_{,}c^{\prime}}=\eta_{c_{,}c^{\prime}}\exp\left\{\!
2i\delta_{c_{,} c^{\prime}}\!\right\}\!.
\end{equation}
This is the traditional representation for the many-channel
systems.\,\,In the second representation, we deal with the uncoupled
eigenchannels, each of them is determined by the eigenphase shift
$\delta_{\nu}$ or the $S_{\nu}$-matrix
\begin{equation}\label{eq:006}
S_{\nu}=\exp\left\{2i\delta_{\nu}\right\}\!,
\end{equation}

\begin{figure}%
\includegraphics[width=\column]{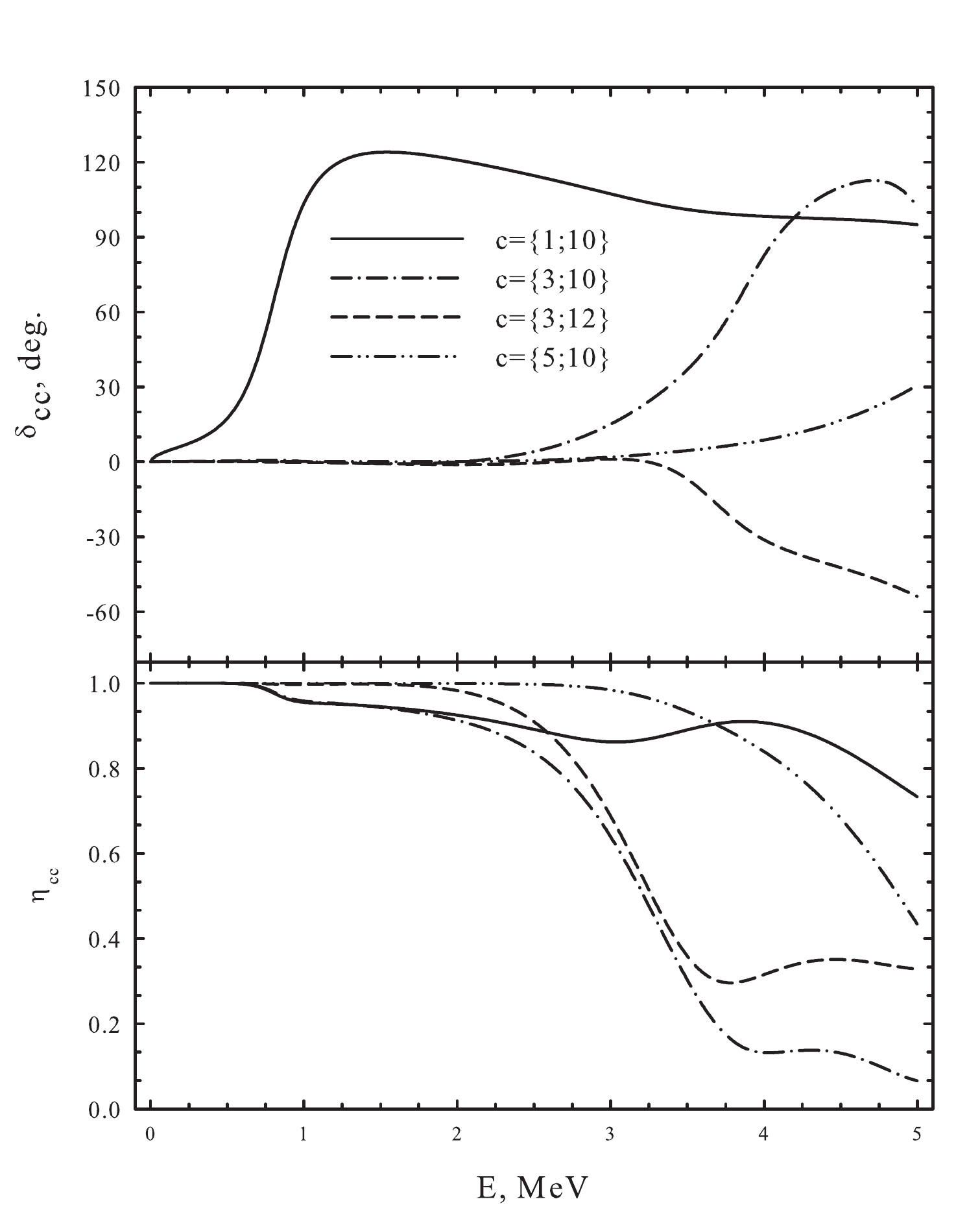}
\vskip-3mm\caption{ Phase shifts and inelastic parameters for four
dominant channels in the $1^{-}$ state of $^{6}$He
}\label{Fig:PhasesEtas1MV2}
\end{figure}

\begin{figure}%
\vskip1mm
\includegraphics[width=8.0cm]{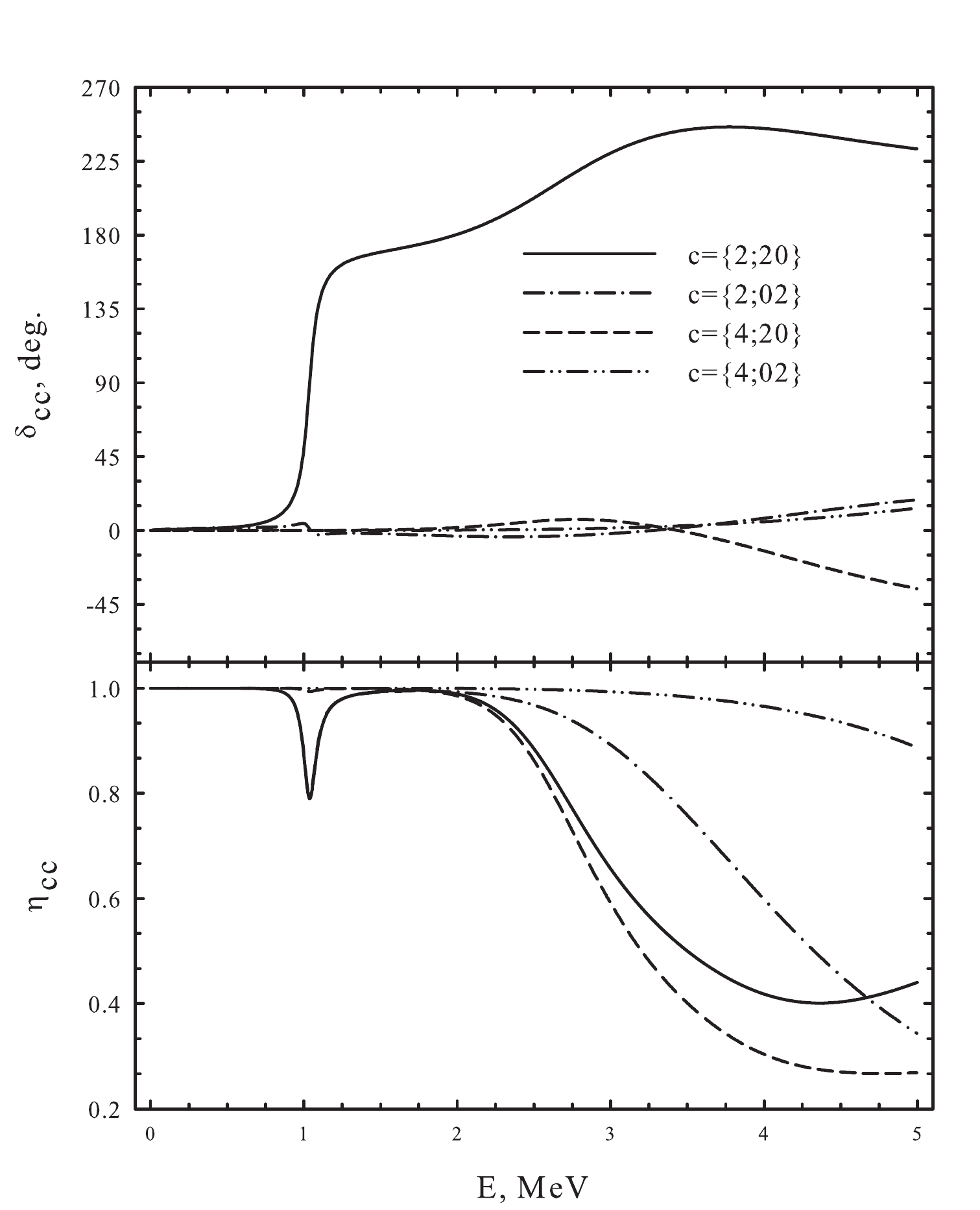}
\vskip-3mm\caption{ Phase shifts and inelastic parameters of the $2^{+}$
state as a function of the energy }\label{Fig:PhasesEtas2PV2}
\end{figure}

\noindent where $\nu$ ($=1$, 2, ..., $N_{\rm ch}$) numerates
eigenchannels.\,\,The relation between the original $\left\Vert
S_{c_{,}c^{\prime}}\right\Vert $ and diagonal $\left\Vert
S_{\nu}\right\Vert $ forms of the $S$- matrix is
\begin{equation}\label{eq:007}
S_{c_{,}c^{\prime}}=\sum_{\nu}U_{\nu}^{c}S_{\nu}U_{\nu}^{c^{\prime}},
\end{equation}
where $\left\Vert U_{\nu}^{c}\right\Vert $ is an orthogonal
matrix.\,\,The energy and the width of a resonance state are
determined from the
eigenphase shift:
\[
\left.  \frac{d^{2}\delta_{\nu}}{dE^{2}}\right\vert
_{E=E_{R}}=0,\quad \Gamma=2\left[  \left.
\frac{d\delta_{\nu}}{dE}\right\vert _{E=E_{R}}\right] ^{-1}\!.
\]
Our experience says (see, e.g., \cite{2001PhRvC..63f4604V,
 2010PPN....41..716N}) that the resonance states of a three-cluster
system usually manifest themselves in one specific eigenchannel.

We now turn our attention to the scattering parameters
($\delta_{c_{,}c^{\prime}} $, $\eta_{c_{,}c^{\prime}}$) and the
parameters of resonances ($E$, $\Gamma$).\,\,In Fig.
\ref{Fig:PhasesEtas1MV2}, we display the phase shifts and the
inelastic parameters for the diagonal elements of the $S$-matrix for
the $J^{\pi} =1^{-}$ state.\,\,These parameters are obtained with
the VP N2 ($m=0.615$).\,\,One can see a resonance behavior of the
phase shift connected with the first channel
$c_{1}=\left\{K=1,l_{1}=1,l_{2}=0\right\}$ at the energy around 1
MeV.\,\,In this energy region, the inelastic parameters show a small
probability of the inelastic processes.\,\,The second resonance
state is observed around 4 MeV and mainly connected with the second
channel $c_{2}=\left\{K=3,l_{1}=1,l_{2}=0\right\}$.

In Fig.\,\,\ref{Fig:PhasesEtas2PV2}, we show the phase shifts and
the inelastic parameters for the $2^{+}$ scattering state.

The well-known $2^{+}$ resonance state realizes itself as a
classical Breit--Wigner resonance state in a few-channel
system.\,\,It displays itself in the phase shift associated with the
channel \{2,20\}.\,\,In the phase shift connected with the channel
\{2,02\}, it is realized as a shadow resonance state.\,\,The
inelastic parameters connected with both dominant channels have a
minimum at the resonance energy.\,\,The second wide $2^{+}$
resonance state involves more channels.\,\,Around this resonance,
the strong rearrangements of compound nuclei are observed, which is
confirmed by the behavior of the inelastic parameters $\eta_{cc}$.

There are many contradictory results in the literature about the
$1^{-}$ resonance state.\,\,The complex scaling method (CSM)
\cite{1993PhRvC..48..165C,
  1995PhRvC..51.2488A} and the hyperspherical
harmonics method \cite{1997PhRvC..55..577D} considering $^{4}$He as
a structureless particle do not confirm the existence of the
state.\,\,In Ref.\,\,\cite{2003PhRvC..68c4313A} within a new version
of CSM, the $1^{-}$ state was obtained as a very broad resonance
($E=3.0$~MeV, $\Gamma= 31.2$~MeV), which is hard to be detected
experimentally.\,\,In our model, we obtain the narrow $1^{-}$ state
for the Volkov N1 and N2 potentials, and a wide resonance state for
the Minnesota potential.\,\,Let us consider the evolution of the
state by increasing the space of hyperspherical harmonics.\,\,We
start our calculations with the simple case of $K_{\max} =
1$.\,\,Then we use $K_{\max} = 3$, 5 and so on up to
$K_{\max}=13$.\,\,The results of these calculations, performed with
the Volkov N2 potential ($m=0.615$) and the Minnesota potential
($u=0.98$), are displayed in Table~\ref{Tab:Coverg1Mreson}.

 One sees that there is no $1^{-}$ resonance state, when we use the hyperspherical
harmonics with $K_{\max} = 1$. When we involve all hyperspherical
harmonics with $1\leq K\leq3$, $1\leq K\leq5$, and $1\leq K\leq7$,
we obtain a broad resonance state, where the total width of the
resonance is larger than its energy.\,\,We assume that it is very
difficult to reveal such broad resonance state by other methods.
Starting from $K_{\max} = 9$, the $1^{-}$ resonance calculated with
the Volkov potential turns out to be a narrow one.\,\,However, for
the Minnesota potential, the width of the $1^{-}$ resonance remains
larger than its energy.\,\,The results presented in Table
\ref{Tab:Coverg1Mreson} can explain why the $1^{-}$ resonance state
was not observed in some calculations, for instance, in the CSM.
When we use a part of the total Hilbert space $K_{\max} =3$, we
impose a restriction on the possible values of partial orbital
momenta $l_{1},l_{2}$: $1\leq l_{1}+l_{2}\leq3$.\,\,We assume that
this restriction is crucial for the formation of the $1^{-}$
resonance state.\,\,By enlarging the space of hyperspherical
harmonics with $K_{\max}\geq 9$, we use a larger number of channels
with $1\leq l_{1}+l_{2}\leq K_{\max}$.\,\,This allows us to describe
more correctly the internal part of the resonance wave function and
its asymptotic part responsible for the decay of the resonance
state.\,\,By closing the discussion about the $1^{-}$ resonance
state, we note that, within a similar three-cluster model
\cite{2009PhRvC..80d4310D}, which combines the hyperspherical
harmonics formalism with the generator coordinate method, the broad
$1^{-}$ resonance state was obtained (the energy and the width of
the resonance state are not specified).\,\,This result, which was
obtained with the Minnesota potential, partially coincides with our
results (see Table \ref{Tab:6HeResonVolkM}).\,\,Indeed, with the
Minnesota potential, the width of the $1^{-}$ resonance is much
larger than that of the resonance calculated with the Volkov
potential.

\begin{table}[b!]
\noindent\caption{ Convergence of the energy\\ and the width (both in MeV)
of the \boldmath$1^-$ resonance state }\vskip3mm\tabcolsep4.2pt

\noindent{\footnotesize\begin{tabular}{|c|c|c|c|c|c|c|c|c|}
  \hline
\multicolumn{1}{|c}{\rule{0pt}{5mm}} &
\multicolumn{1}{|c}{$K_{\max}$ } & \multicolumn{1}{|c}{1} &
\multicolumn{1}{|c}{3 } & \multicolumn{1}{|c}{5} &
\multicolumn{1}{|c}{7 } & \multicolumn{1}{|c}{9} &
\multicolumn{1}{|c}{11 } &
\multicolumn{1}{|c|}{13 } \\[2mm]
\hline
\rule{0pt}{5mm}VP  & $E$ & -- & 1.697 & 1.527 & 1.292 & 1.043 & 0.911 & 0.815 \\
  N2 & $\Gamma$ & -- & 2.996 & 2.022 & 1.311 & 0.788 & 0.585 & 0.460 \\
  MP & $E$ & -- & 2.274 & 2.250 & 2.142 & 1.945 & 1.871 & 1.775 \\
  &{$\Gamma$} & -- & 6.399 & 4.665 & 3.577 & 2.476 &
2.116 & 1.780 \\[2mm]
  \hline
\end{tabular}\label{Tab:Coverg1Mreson}}
\end{table}

\begin{table}[b!]
\vskip3mm \noindent\caption{ Contribution of different states\\ with
the total orbital momentum \boldmath$L$ and the total\\ spin $S$ to
the energy and the width\\ of the $1^-$ resonances in $^6$He
}\vskip3mm\tabcolsep9.3pt

\noindent{\footnotesize\begin{tabular}{|c|c|c|c|}
  \hline
\multicolumn{1}{|c}{\rule{0pt}{5mm}$(LS)$} &
\multicolumn{1}{|c}{(10) } & \multicolumn{1}{|c}{$(10)+(11)$ } &
\multicolumn{1}{|c|}{$(10)+(11)+(21)$ } \\[2mm]
\hline
\rule{0pt}{5mm}$E$, MeV & 1.842 & 1.776 & 1.775 \\
  $\Gamma$, MeV & 2.750 & 1.783 & 1.780 \\
  $E$, MeV & 3.892 & 3.222 & 3.264 \\
  $\Gamma$, MeV & 3.597 & 3.698 & 3.635 \\[2mm]
  \hline
\end{tabular}}\label{Tab:LScontrib1M}
\end{table}

\begin{table}[b!]
\vspace*{-1mm} \noindent\caption{ Spectrum of resonance states\\ in
\boldmath$^6$He obtained with different $NN$ potentials.\\ Energy
$E$ and width $\Gamma$ are in MeV }\vskip3mm\tabcolsep3.9pt

\noindent{\footnotesize\begin{tabular}{|c|c|c|c|c|c|c|c|c|c|}
  \hline
 & \multicolumn{2}{|c}{\rule{0pt}{5mm}VP N1} & \multicolumn{2}{|c|}{VP N2} &
\multicolumn{2}{|c|}{VP N2} & \multicolumn{3}{|c|}{MP}\\[2mm]
\cline{2-10} & \multicolumn{2}{|c}{\rule{0pt}{5mm}$m=0.60$} &
\multicolumn{2}{|c}{$m=0.60$} &
\multicolumn{2}{|c|}{$m=0.615$} & \multicolumn{3}{|c|}{$u=0.98$}\\[2mm]
  \hline
 \rule{0pt}{5mm}$J^{\pi}$ & $E$ & $\Gamma$ & $E$ & $\Gamma$ & $E$ & $\Gamma$ & $E$ & $\Gamma$ & $J^{\pi}$ \\
  $0^{+}$ & 0.0 & -- & 0.0 & -- & 0.0 & -- & 0.0 & -- & $0^{+}$ \\
  $1^{-}$ & 1.80 & 0.65 & 2.02 & 0.43 & 1.79 & 0.47 & 1.03 & 1 keV 
 & $2^{+}$ \\
  $2^{+}$ & 2.01 & 0.15 & 2.10 & 0.05 & 2.01 & 0.11 & 2.32 & 1.63 & $1^{-}$ \\
  $0^{+}$ & 2.25 & 1.98 & 2.506 & 1.71 & 2.34 & 1.87 & 3.10 & 1.04 & $2^{+}$ \\
  $2^{+}$ & 3.54 & 1.96 & 3.64 & 1.80 & 3.56 & 1.74 & 3.12 & 1.29 & $2^{-}$ \\
  $1^{-}$ & 4.21 & 2.41 & 4.44 & 2.00 & 4.26 & 2.18 & 3.59 & 1.87 & $1^{+}$ \\
  $2^{-}$ & 5.09 & 4.97 & 7.03 & 5.25 & 5.30 & 5.40 & 3.65 & 3.34 & $0^{+}$ \\
  $1^{+}$ & 5.13 & 2.59 & 5.37 & 2.24 & 5.20 & 2.50 & 3.77 & 3.76 & $1^{-}$ \\
  &  &  &  &  &  &  & 5.20 & 2.51 & $1^{+}$ \\[2mm]
  \hline
\end{tabular}\label{Tab:6HeResonVolkM}}
\end{table}

\begin{figure}%
\vskip1mm
\includegraphics[width=\column]{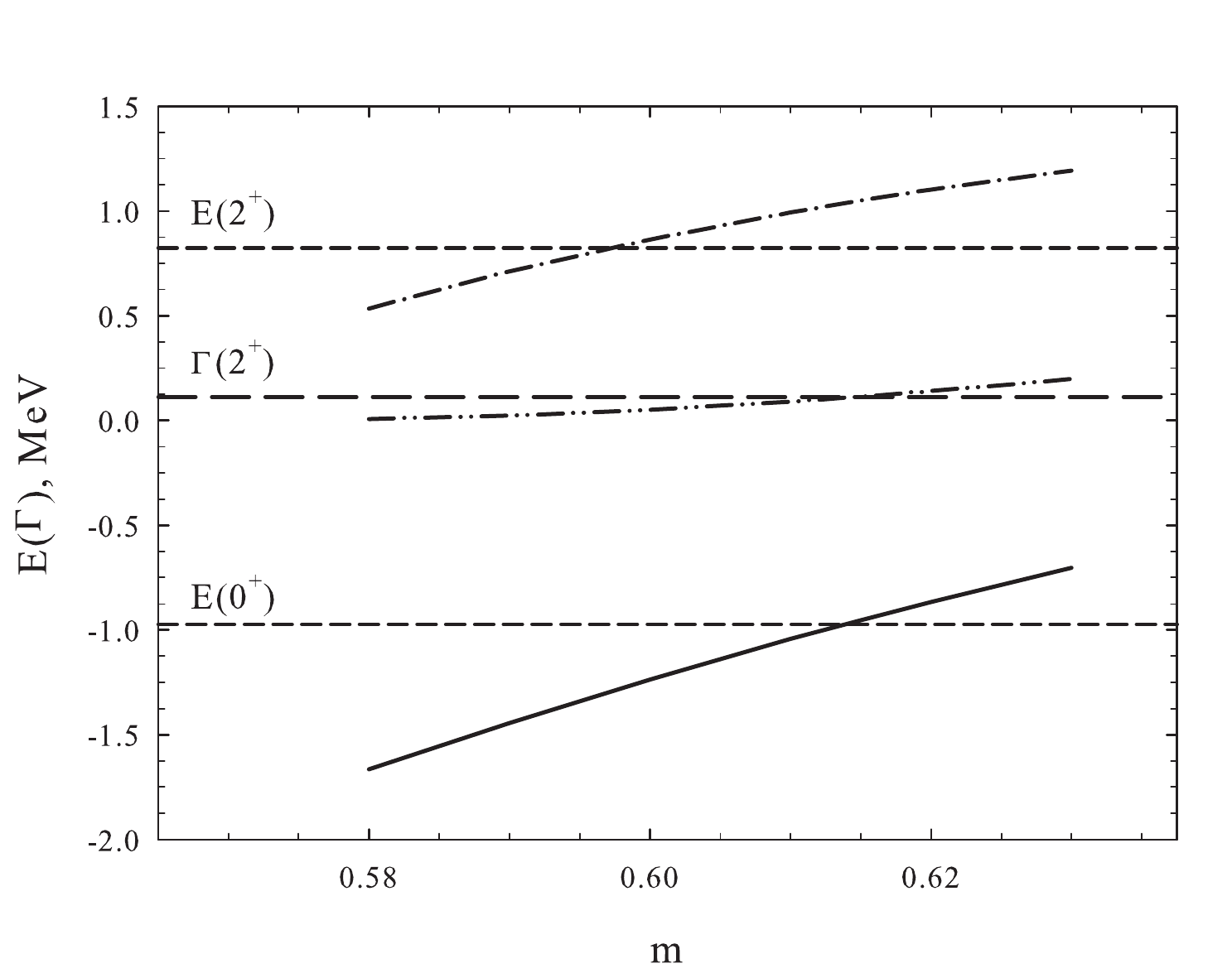}
\vskip-3mm\caption{ Energy of the ground state (solid line), energy
(dash-dotted line) and width (dash-dot-dotted line) of $2^{+}$
resonance state as functions of the Majorana parameter $m$ (VP
N2).\,\,Experimental values of these quantities are indicated by
dashed line }\label{Fig:EnergyvsM}
\end{figure}

In Table \ref{Tab:LScontrib1M}, we show how the contributions of
different values of the total orbital momentum $L$, total spin $S,$
and spin-orbital components of the Minnesota potential affect the
parameters of the first and second $1^{-}$ resonance states.\,\,The
results presented in Table \ref{Tab:LScontrib1M} can shed some more
light on the problem of existence of the $1^{-}$ resonance state in
$^{6}$He.\,\,If we restrict ourselves with the total spin $S=0$ and
thus switch-off the spin-orbital components of the Minnesota
potential, we obtain a very wide resonance state.\,\,By adding the
component $( LS) =( 11)$ to the total wave function, we slightly
decrease the energy of the first $1^{-}$ resonance state and
significantly reduce its width.\,\,The next component $( LS) =( 21)$
does not change the parameters of the resonance.\,\,Thus, the first
$1^{-}$ resonance state remains a wide resonance state, which, we
believe, is difficult to be determined by the complex scaling method
or other methods.\,\,Note that the second resonance state became
wider, as we increase the space of $( LS)  $ components.\,\,Indeed,
the energy of the state is decreased, while the total width remains
the same.

In Table \ref{Tab:6HeResonVolkM}, we display the spectrum of
resonance states in $^{6}$He obtained with different $NN$
potentials.\,\,The energy of resonances is reckoned from the
ground-state energy.\,\,The total angular momenta of resonance
states for the Volkov potential are indicated on left-hand side of
the table, while, for the Minnesota potential, they are shown on the
right-hand side of the table.\,\,We note that the Volkov potentials
N1 and N2 yield a similar set of resonance states.\,\,Moreover, the
Volkov potential N1 and N2 generate the narrow $1^{-}$ resonance
state.\,\,The resonance width is between 0.473 and 0.647 MeV, and
its energy is less than the energy of the well-known $2^{+}$
resonance state.\,\,The latter is obtained with an energy of
2.012--2.096 MeV, which is $\approx 300$ keV larger than the
experimental energy.\,\,Contrary to the Volkov potential, the
Minnesota potential creates a very narrow $2^{+}$ resonance state
($\Gamma$=1.0 keV) with an energy of 1.033 MeV, which is much less
than the experimental value.\,\,As is seen, the $0^{+}$\ and $2^{-}$
resonance states, calculated with the Volkov N2 potential, are too
broad.\,\,So, it is very difficult to detect experimentally such
short-living states.

\begin{figure}%
\vskip3mm
\includegraphics[width=\column]{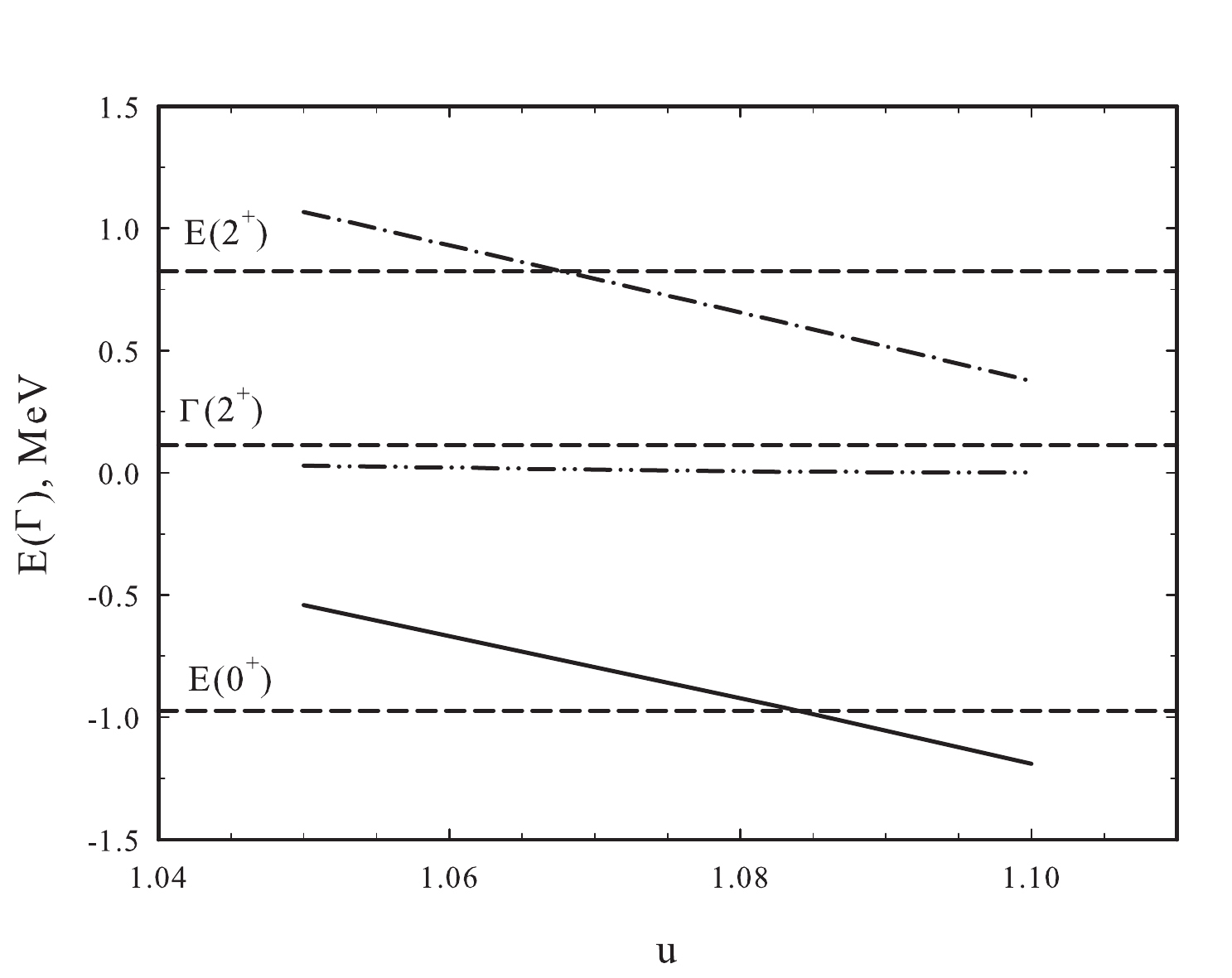}
\vskip-3mm\caption{ Energy of $0^{+}$ bound state, energy and width
of the $\ 2^{+}$ resonance as a function of parameter $u$ of the MP
}\label{Fig:EnergyvsU}
\end{figure}

We found that, with the selected $NN$ potentials, it is impossible
to obtain simultaneously the correct values of energies of the
$0^{+}$ bound state and the $2^{+}$ resonance state (with respect to
the energy of the three-cluster threshold).\,\,This statement is
valid for the present model.\,\,However, we believe that they can be
obtained in other alternative models.\,\,This statement is
demonstrated in Fig.\,\,\ref{Fig:EnergyvsM}, where we display the
the dependence of energy of the bound and $2^{+}$ resonance states
on the Majorana parameter of the VP.\,\,Note that this parameter is
used in many microscopic calculations as a variational
parameter.\,\,In the cluster model or the resonating group method,
the Majorana parameter is selected to reproduce the main properties
of two-cluster subsystems of the compound three-cluster
system.\,\,One can also see that the relative position of the
$0^{+}$ and $2^{+}$ states slightly changes with a variation of the
Majorana parameter.\,\,In addition, the optimal parameter $m$ for
the $0^{+}$ state (where the energy coincides with the experimental
one) differs from the optimal value $m$ for the $2^{+}$ resonance
state and its width.\,\,A similar picture is observed for the
Minnesota potential (see Fig.\,\,\ref{Fig:EnergyvsU}).\,\,The
$2^{+}$ resonance state width obtained with the MP is very small
with respect to the experimental value and is slightly changing in
the displayed range of the parameter $u$.


Now we determine the dominant channels for the decay of
resonances.\,\,For this aim, we calculate the partial widths of
resonance states.\,\,In Table \ref{Tab:6HePartWidthsV2}, we display
the partial widths for resonances in $^{6}$He calculated with the
Volkov potential N2.\,\,Below each partial width $\Gamma_{\alpha}$
($\alpha=1,2, ...$), we also indicate the quantum numbers of
channels $c_{\alpha}=$($K;l_{1}l_{2};LS$), into which a resonance
state decays.\,\,One notices that we display only three partial
widths.\,\,We show only dominant decay channels for a resonance
state.\,\,The total contribution of these channels to the total
widths is at least 96\%.\,\,Thus, the contribution of the omitted
channels is very small.

Let us analyze the quantum numbers of the dominant decay
channels.\,\,One can see, for example, that the first $2^{+}$
resonance state decays mainly through the channel with hypermomentum
$K=2$ and with the zero value of orbital momentum of two valence
neutrons.\,\,Two dominant channels represent the decay of the second
$2^{+}$ resonance state.\,\,The first of them exhausts approximately
73\% of the total width, while the second channel exhausts
approximately 25\%.\,\,One also observes that all resonance states
displayed in Table \ref{Tab:6HePartWidthsV2} have one or at most two
dominant channels.
 Similar results are obtained with other NN-po\-tentials.


\begin{table}[b]
\noindent\caption{ Total \boldmath$\Gamma$ and partial $\Gamma_i$
($i=1$, 2, 3)\\ widths for resonance states in $^6$He calculated\\
with the VP N2, $m=0.615$ }\vskip3mm\tabcolsep3.8pt

\noindent{\footnotesize\begin{tabular}{|c|c|c|c|c|c|}
  \hline
\multicolumn{1}{|c}{\rule{0pt}{5mm}$J^{\pi}$} &
\multicolumn{1}{|c}{$E$, MeV } & \multicolumn{1}{|c}{$\Gamma$, MeV }
& \multicolumn{1}{|c}{$\Gamma_{1}/c_{1}$ } &
\multicolumn{1}{|c}{$\Gamma_{2}/c_{2}$ } &
\multicolumn{1}{|c|}{$\Gamma_{3}/c_{3}$ } \\[2mm]
\hline \rule{0pt}{5mm}$2^{+}$ & 1.033 & 0.107 & 0.096 & 0.011 & 0.0003 \\
  &  & ($K;l_{1}l_{2};LS$) & (2;20;20) & (2;02;20) & (4;20;20) \\
  $1^{-}$ & 0.808 & 0.473 & 0.466 & 0.006 & 0.0006 \\
  &  & ($K;l_{1}l_{2};LS$) & (1;10;10) & (3;10;10) & (3;12;10) \\
  $0^{+}$ & 1.364 & 1.869 & 1.028 & 0.830 & \\
  &  &  & (0;00;00)& (2;00;00) & \\
  $2^{+}$ & 2.581 & 1.735 & 1.274 & 0.023 & 0.429 \\
  &  & ($K;l_{1}l_{2};LS$) & (2;20;20) & (2;02;20) & (4;20;20) \\
  $2^{-}$ & 4.317 & 5.399 & 4.930 & 0.001 & 0.442 \\
  &  &  & (3;12;20) & (3;21;20) & (3;01;20) \\
  $1^{+}$ & 4.170 & 2.370 & 2.356 & 0.013 & 0.012 \\
  &  &  & (4;22;10) & (4;31;10) & (4;13;10) \\
  $1^{-}$ & 3.284 & 2.184 & 0.183 & 1.186 &0.690 \\
  &  &  & (1;10;10) & (3;10;10) & (3;12;10) \\[2mm]
  \hline
\end{tabular}}\label{Tab:6HePartWidthsV2}
\end{table}

Here, we present the correlation functions in the coordinate and
momentum spaces.\,\,The definitions of these quantities are given in
\cite{2007JPhG...34.1955B}.\,\,The correlation functions allow one
to determine the most probable geometrical configuration of a
three-cluster system.\,\,This can be done both for the bound and
resonance states.\,\,In Ref.\,\,\cite{2007JPhG...34.1955B}, we
formulated an algorithm of selecting the ``true'' resonance function
among a large number of wave functions of the many-channel system.
The algorithm allows us to analyze only one function or only one
correlation function connected with the selected resonance state.

\begin{figure}%
\vskip1mm
\includegraphics[width=7.7cm]{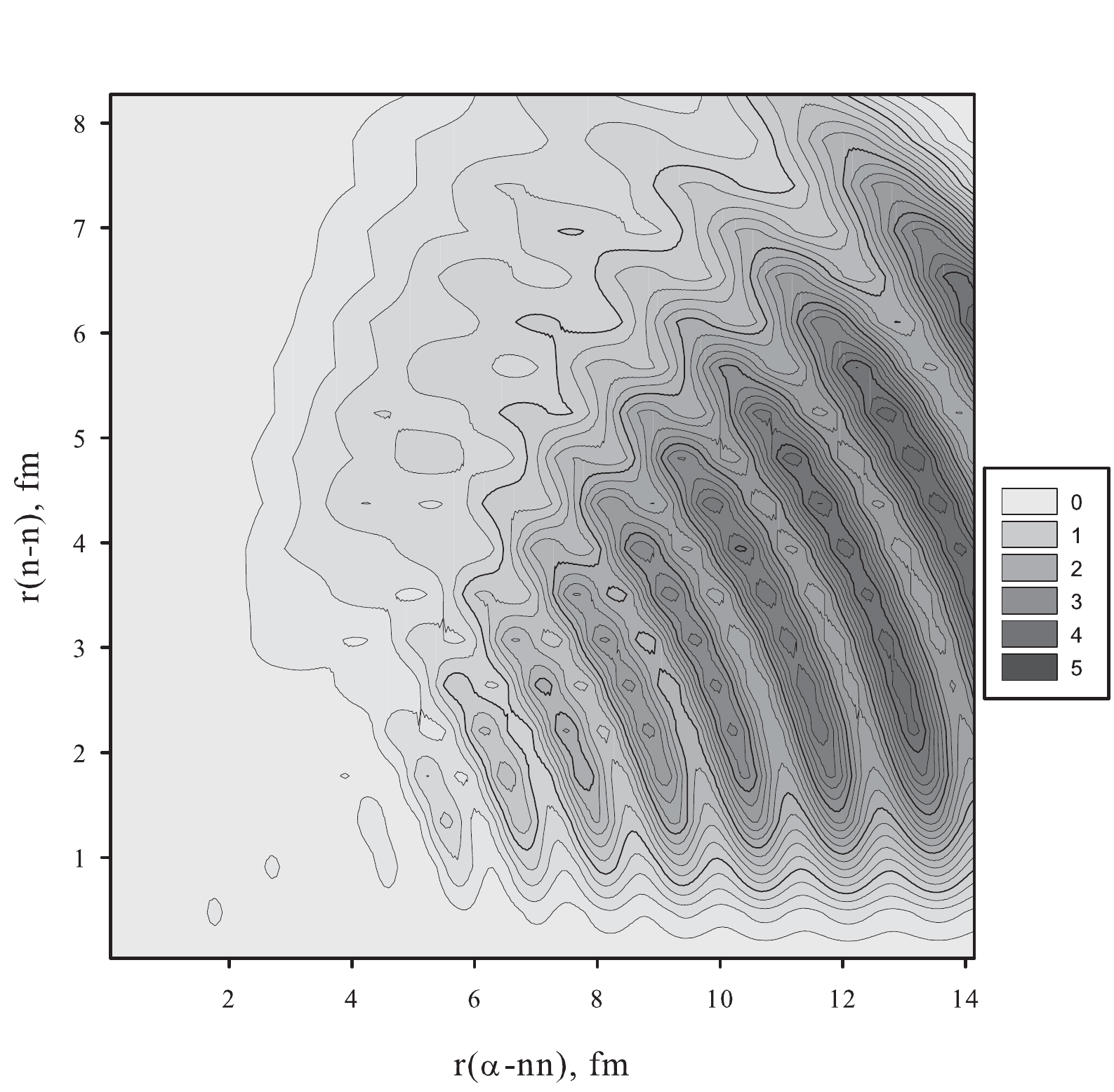}
\vskip-3mm\caption{ Correlation function for the $1^{-}$ resonance
state calculated with the MP }\label{Fig:CorrFun1MresonCSMP}
\end{figure}

\begin{figure}
\vskip3mm
\includegraphics[width=\column]{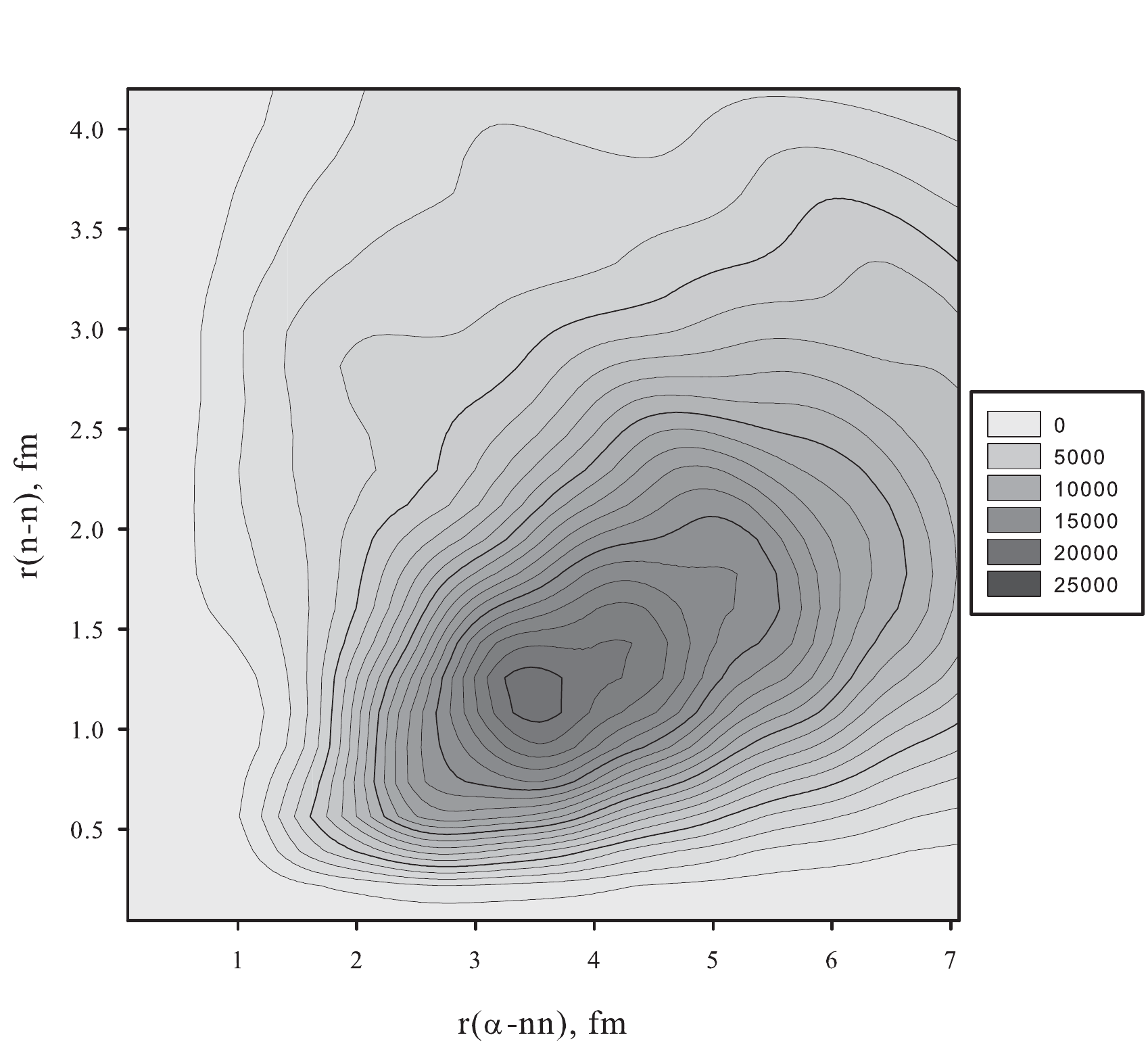}
\vskip-3mm\caption{ Correlation function for the first $2^{+}$
resonance state calculated with the Minnesota potential
}\label{Fig:CorrFun2Preson1CSMP}
\end{figure}

In Fig.\,\,\ref{Fig:CorrFun1MresonCSMP}, we display the correlation
function for the first $1^{-}$ resonance state.\,\,Figure
\ref{Fig:CorrFun2Preson1CSMP} shows the correlation function for the
first $2^{+}$ resonance state.

\begin{figure}%
\vskip1mm
\includegraphics[width=\column]{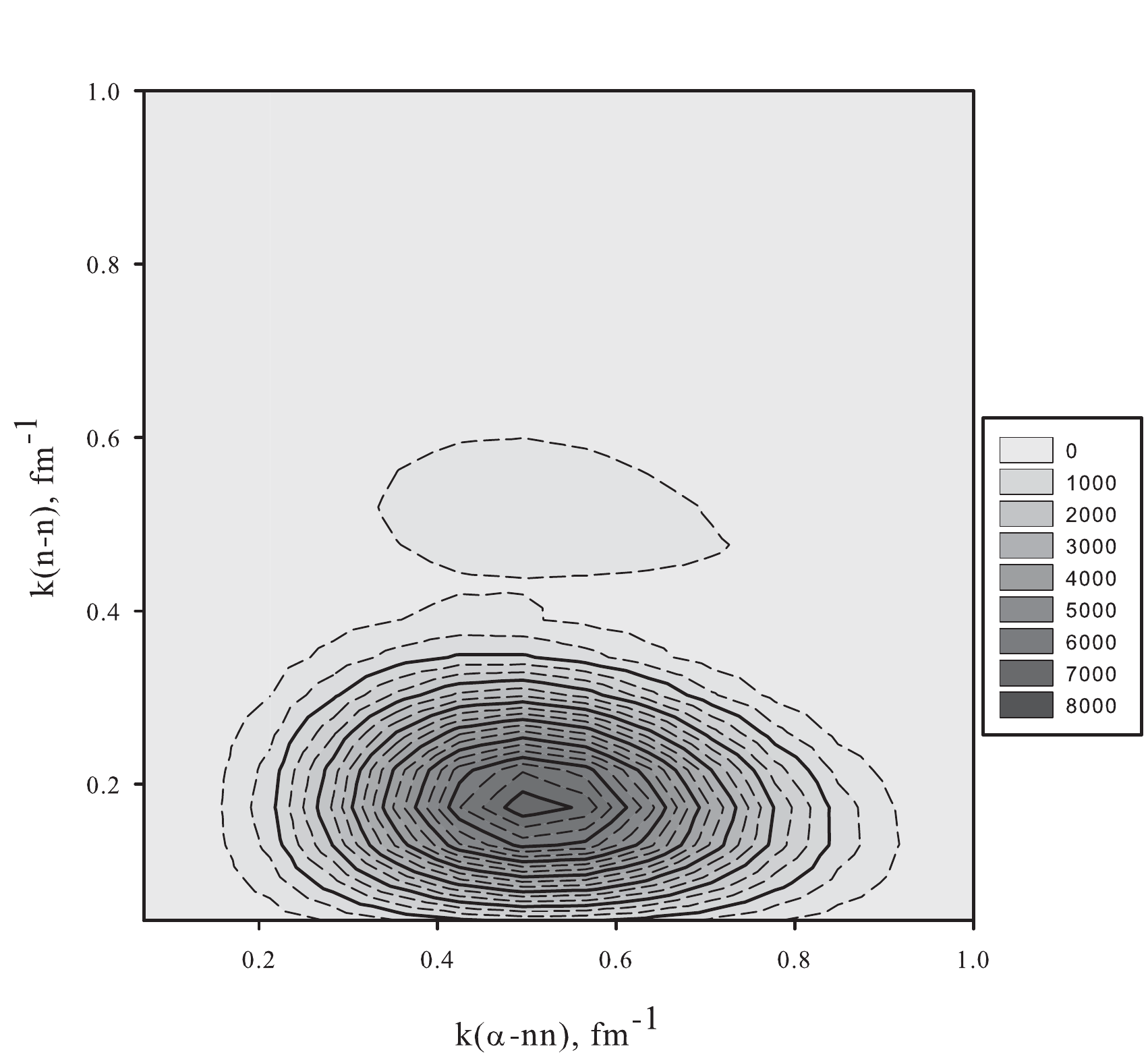}
\vskip-3mm\caption{ Correlation function for the $1^{-}$ resonance
state calculated with the MP in the momentum space
}\label{Fig:CorrFun1MresonMSMP}
\end{figure}

One sees the common feature of these figures: a small distance
between valence neutrons $r\left(n-n\right)$ and a large distance
between the alpha-particle and the center of masses of the
two-neutron subsystem $r\left(\alpha-nn\right)$.\,\,Thus, the
dominant configuration of $^{6}$He in these two resonances is an
obtuse triangle.\,\,The correlation function of the very narrow
$2^{+}$ resonance state behaves itself as the correlation functions
of a bound state.\,\,This is the common feature of very narrow
resonance states, which have the wave function very close to wave
functions of quasibound states with the same energy.

\begin{figure}%
\vskip1mm
\includegraphics[width=\column]{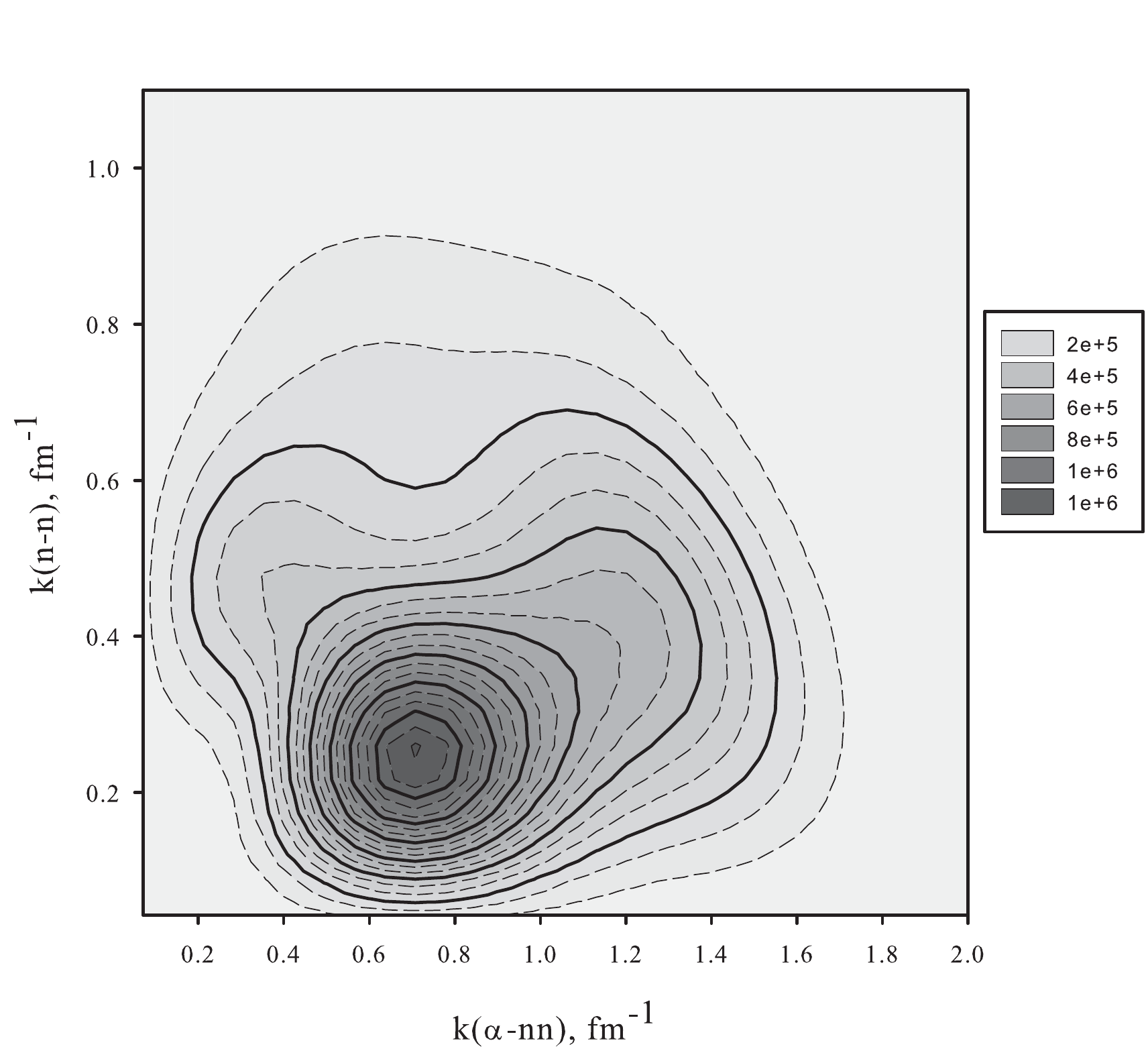}
\vskip-3mm\caption{ Correlation function determined in the momentum
space with the MP for the first $2^{+}$ resonance state
}\label{Fig:CorrFun2Preson1MSMP}
\end{figure}

The additional information about the three-cluster resonance state
can be obtained by analyzing its correlation function in the
momentum space.\,\,In Figs.\,\,\ref{Fig:CorrFun1MresonMSMP} and
\ref{Fig:CorrFun2Preson1MSMP}, we display the correlation functions
of the $1^{-}$ and $2^{+}$ resonance states, respectively.\,\,Both
pictures were obtained with the Minnesota potential.\,\,The main
feature of the correlation functions is that the alpha particle has
a larger relative momentum $k\left(\alpha-nn\right)$ than the
relative momentum of two valence neutrons $k\left(n-n\right)$.

To get a more information about the three-cluster resonance, we
consider the decay cross section for a resonance state.\,\,Having
calculated elements of the scattering $S$-matrix, we can determine
the cross sections of the elastic and rearrangement processes, which
proceed in the three-cluster continuum.\,\,Some examples of cross
sections, which are determined by one hyperspherical harmonic, were
demonstrated in~\cite{2001PhRvC..63f4604V}.

It is well known (see, e.g., Refs.\,\,\cite{1971RSPTA.270..197G,
 1972AnPhy..74..324N, 1977TMPh..31..308B,
 1980CzJPh..30.1090J}) that the $3\Rightarrow 3$ scattering
amplitude depends on 5 parameters ($\mathbf{k}_{0}$,
$\mathbf{q}_{0}$) connected with the incoming channel and 5
parameters ($\mathbf{k}$, $\mathbf{q}$) attributed to the exit
channel.\,\,They are momenta associated with the Jacobi
vectors.\,\,In what follows, we determine the vector $\mathbf{k}$ as
the momentum of the relative motion of a selected pair of clusters
and the vector $\mathbf{q}$  as the momentum of motion of the third
cluster with respect to the center of mass of the selected pair of
clusters.\,\,These vectors satisfy the relation
\begin{equation}\label{eq:Energy}
E=\frac{\hbar^{2}}{2m}\mathbf{k}_{0}^{2}+\frac{\hbar^{2}}{2m}\mathbf{q}
_{0}^{2}=\frac{\hbar^{2}}{2m}\mathbf{k}^{2}+\frac{\hbar^{2}}{2m}\mathbf{q}
^{2},
\end{equation}
where $E$ is the total energy of the three-cluster system.\,\,Thus,
not all of the parameters are independent.\,\,There are other
relations, which reduce the number of independent parameters, if one
needs to determine the $3\Rightarrow 3$ scattering amplitude.\,\,We
do not dwell on these relations.

In our case (in the representation of the hyperspherical harmonics), the
amplitude is defined as
\[f\left( \mathbf{k}_{0},\mathbf{q}_{0};\mathbf{k},\mathbf{q}\right)
=\sum_{K,l_{1},l_{2}}\sum_{\widetilde{K},\widetilde{l}_{1},\widetilde{l}_{2}
}\sum_{M}\chi_{K}^{\left(l_{1},l_{2}\right)}\left(\theta_{0}\right)\times\]\vspace*{-7mm}
\[\times\left(  \delta_{K,l_{1},l_{2};\widetilde{K},\widetilde{l}%
_{1},\widetilde{l}_{2}}-S_{K,l_{1},l_{2};\widetilde{K},\widetilde{l}%
_{1},\widetilde{l}_{2}}\right)  \chi_{\widetilde{K}}^{\left(
\widetilde{l}_{1},\widetilde{l}_{2}\right)  }\left(  \theta\right)
\times\]\vspace*{-7mm}
\begin{equation}\label{eq:CS02}
\times\left\{  Y_{l_{1}}\left(  \mathbf{k}_{0}\right)
Y_{l_{2}}\left(\mathbf{q}_{0}\right)  \right\} _{LM}^{\ast}\left\{\!
Y_{\widetilde{l}_{1} }\left( \mathbf{k}\right)
Y_{\widetilde{l}_{2}}\left(  \mathbf{q}\right) \!\right\}  _{LM}\!.
\end{equation}
The five-fold differential cross section describing rearrangements
in the three-cluster continuum is
\begin{equation}
d\sigma\left(
\mathbf{k}_{0},\mathbf{q}_{0};\mathbf{k},\mathbf{q}\right)
\sim\left\vert f\left(  \mathbf{k}_{0},\mathbf{q}_{0};\mathbf{k}
,\mathbf{q}\right)  \right\vert ^{2}d\mathbf{k}d\mathbf{q}.
\label{eq:CS03}
\end{equation}
One may define the amplitude
$f_{c,\widetilde{c}}\left(\mathbf{k}_{0}
,\mathbf{q}_{0};\mathbf{k},\mathbf{q}\right)$ and the corresponding
cross section $d\sigma_{c,\widetilde{c}}\left(
\mathbf{k}_{0},\mathbf{q} _{0};\mathbf{k},\mathbf{q}\right)$ which
determine the transition of the system from the three-cluster
channel $c$ to the channel $\widetilde{c}$.\,\,Let $\left\vert
c\right\rangle $ be the hyperspherical harmonic
\begin{equation}
\left\vert c;\mathbf{k,q}\right\rangle =\left\vert
K,l_{1},l_{2}\right\rangle =\chi_{K}^{\left(  l_{1},l_{2}\right)
}\left(  \theta\right)  \left\{ Y_{l_{1}}\left(  \mathbf{k}\right)
Y_{l_{2}}\left(  \mathbf{q}\right) \right\}  _{LM}. \label{eq:CS04}
\end{equation}\vspace*{-7mm}

\noindent Then
\begin{equation}
f_{c,\widetilde{c}}\left(  \mathbf{k}_{0},\mathbf{q}_{0};\mathbf{k}
,\mathbf{q}\right) \! =\!\left\vert
c;\mathbf{k}_{0},\mathbf{q}_{0}\right\rangle\! \left[
\delta_{c;\widetilde{c}}-S_{c;\widetilde{c}}\right]  \left\langle
\widetilde{c};\mathbf{k,q}\right\vert.\! \label{eq:CS05}
\end{equation}
In view of Eq.\,(\ref{eq:007}) and the orthogonality properties
of the matrix $\left\Vert U_{\nu}^{c}\right\Vert $, we can rewrite
$f_{c,\widetilde {c}}\left(
\mathbf{k}_{0},\mathbf{q}_{0};\mathbf{k},\mathbf{q}\right)$ as
\[f_{c,\widetilde{c}}\left(  \mathbf{k}_{0},\mathbf{q}_{0};\mathbf{k}%
,\mathbf{q}\right) =\]\vspace*{-7mm}
\[
=\left\vert c;\mathbf{k}_{0},\mathbf{q}_{0}\right\rangle
\sum_{\nu}U_{\nu }^{c}\left[  1\!-\!S_{\nu}\right]
U_{\nu}^{\widetilde{c}}\left\langle
\widetilde{c};\mathbf{k,q}\right\vert=\]\vspace*{-7mm}
\begin{equation}\label{eq:CS06}
=\left\vert c;\mathbf{k}_{0},\mathbf{q}_{0}\right\rangle
\sum_{\nu}U_{\nu }^{c}\left[  1-S_{\nu}\right]
U_{\nu}^{\widetilde{c}}\left\langle
\widetilde{c};\mathbf{k,q}\right\vert.
\end{equation}
This equation connects the eigenchannels $\nu$ with original channels
$c$.

It is a very complicated and cumbersome tremendous expression to be
analyzed and depends on too many input parameters.\,\,We have to
reduce the number of these parameters in order to get information in
a simpler way.\,\,If we integrate this many-fold cross section over
the input parameters $\mathbf{k}_{0}$ and $\mathbf{q}_{0},$ we
obtain\vspace*{-3mm}
\[
d\sigma\left(  \mathbf{k},\mathbf{q}\right)  \sim\left(
{\displaystyle\int}
\left\vert f\left(
\mathbf{k}_{0},\mathbf{q}_{0};\mathbf{k},\mathbf{q} \right)
\right\vert ^{2}d\mathbf{k}_{0}d\mathbf{q}_{0}\!\right) d\mathbf{k}
d\mathbf{q}.
\]
Thus, we obtain now the cross section, which is averaged over the
input parameters or the initial conditions.\,\,The new cross section
$d\sigma\left( \mathbf{k},\mathbf{q}\right)$ depends on the
parameters of exit channels and determines the decay of a
three-cluster resonance state.\,\,The expression for the cross
section can be simplified by the integration over the unit vectors
$\widehat{\mathbf{k}}$ and $\widehat{\mathbf{q}}$
\[
d\sigma\left(  k,q\right)  \sim\left(
{\displaystyle\int}
\sigma\left(  \mathbf{k},\mathbf{q}\right)  d\widehat{\mathbf{k}}
d\widehat{\mathbf{q}}\right)  dkdq.
\]

\begin{figure}%
\vskip1mm
\includegraphics[width=\column]{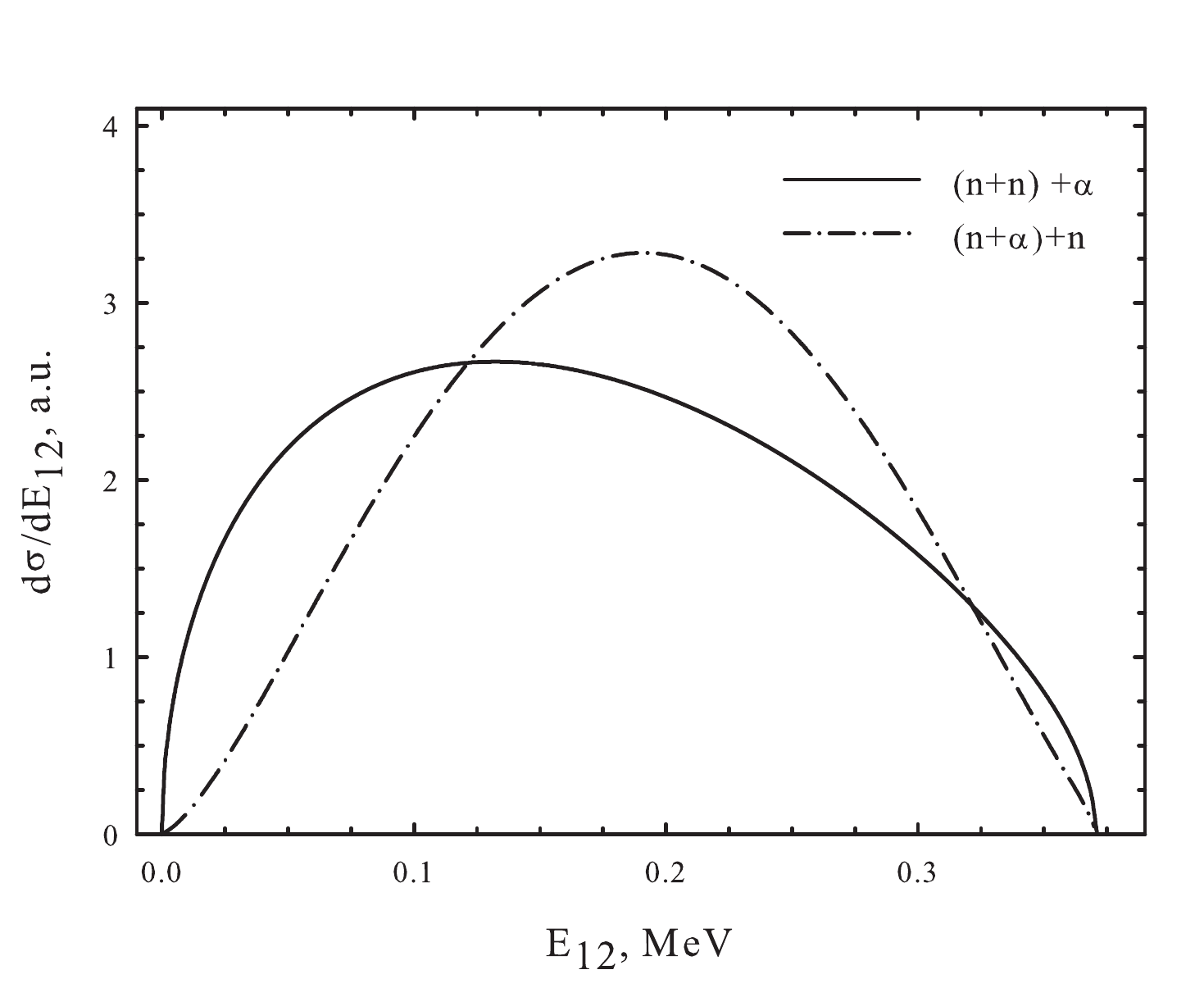}
\vskip-3mm\caption{ Differential cross section for the decay of the
$2^{+}$ resonance state.\,\,$E_{12}$ is the relative energy for a
selected pair of clusters : the $n+n$ subsystem (solid line) and the
$n+\alpha$ subsystem (dot-dashed line).\,\,Results are obtained with
the MP }\label{Fig:CrossSec2PMP}
\end{figure}

\begin{figure}%
\vskip1mm
\includegraphics[width=\column]{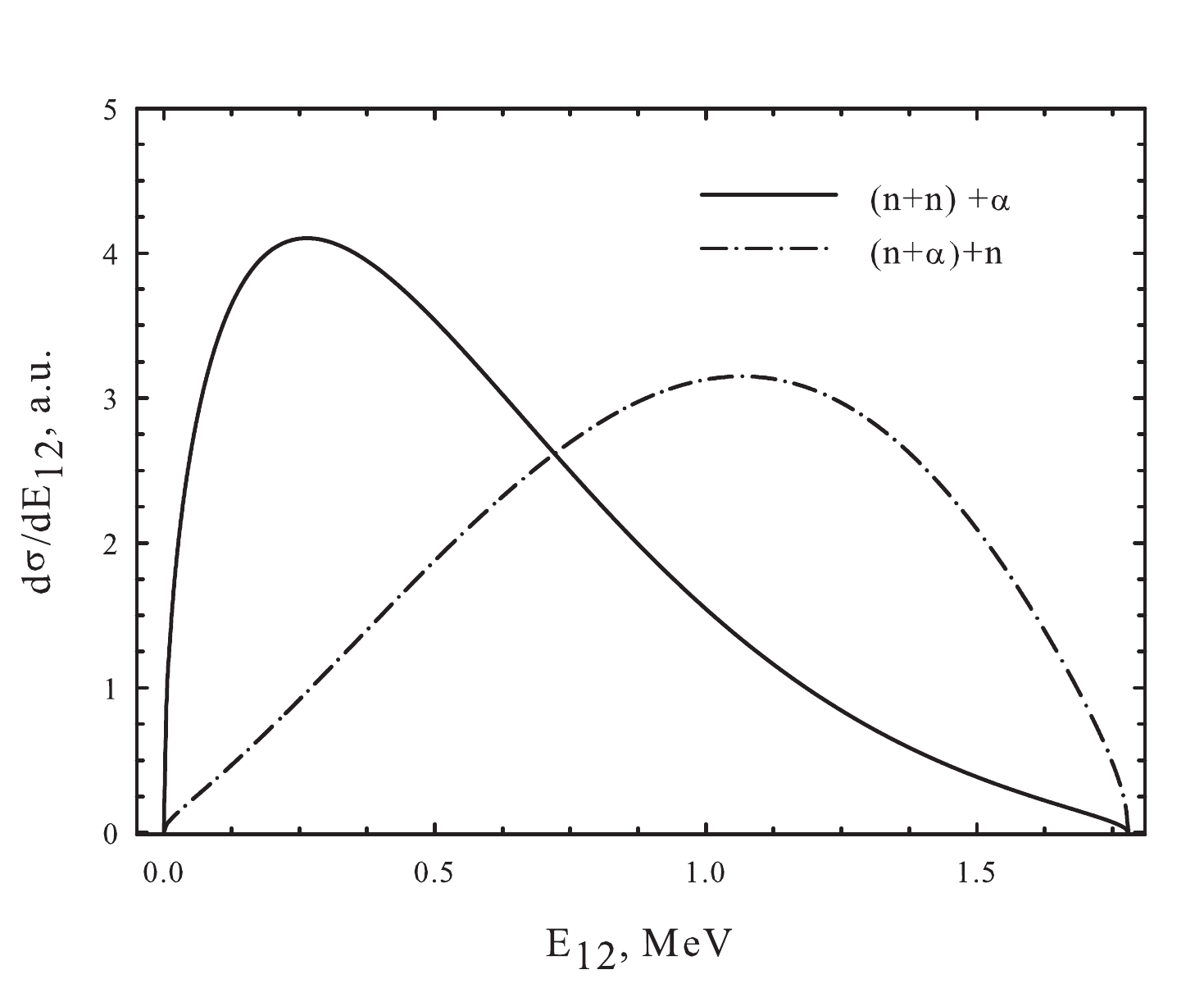}
\vskip-3mm\caption{ One-fold differential cross section, which
determines the dominant energy and channels for the decay of the
$1^{-}$ resonance state.\,\,Results are obtained with the MP
}\label{Fig:CrossSec1MMP}
\end{figure}

\begin{figure}%
\vskip1mm
\includegraphics[width=\column]{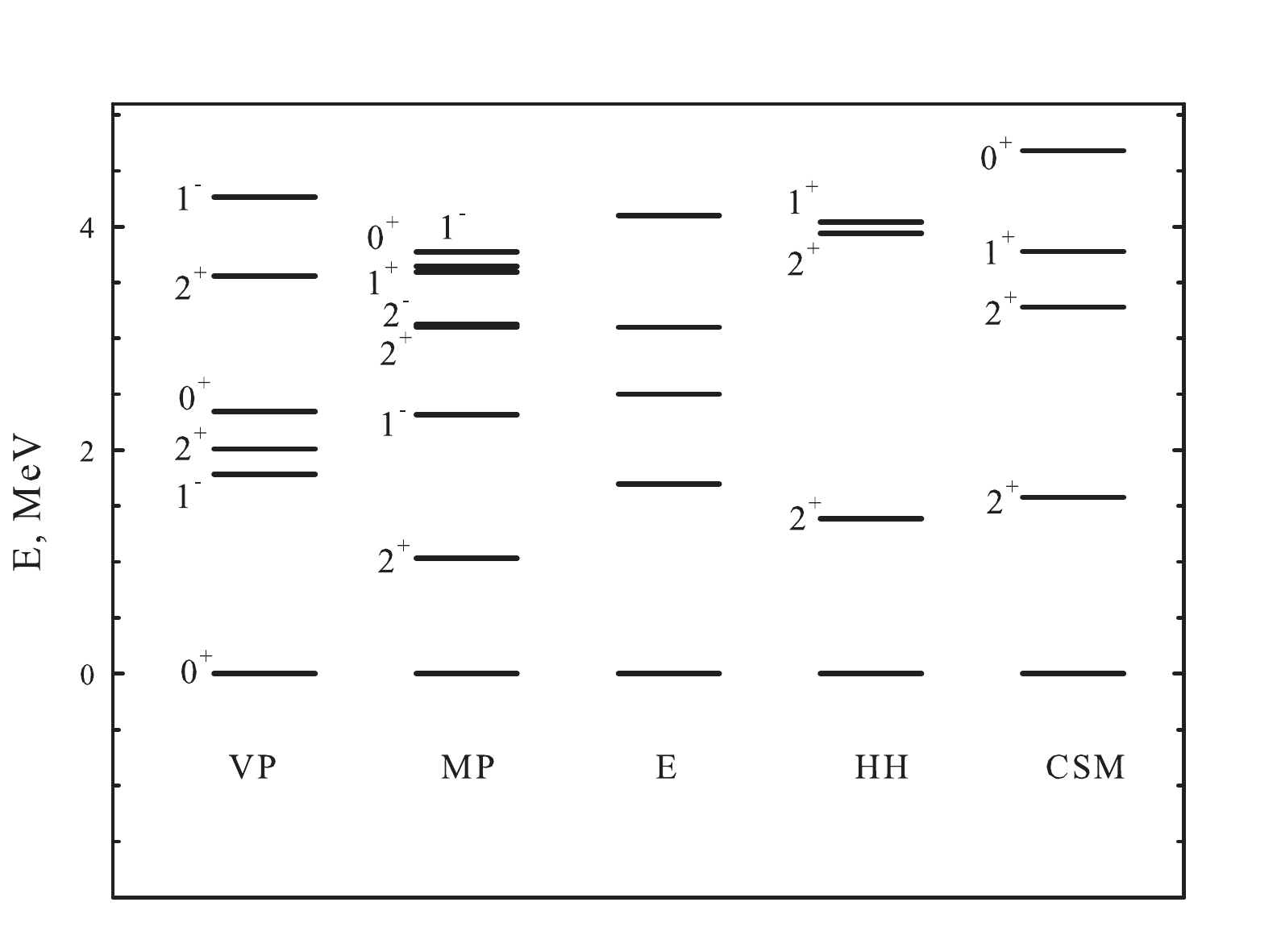}
\vskip-3mm\caption{ Comparison of the experimental ($E$) and
theoretical (VP, MP, HH, CSM, see details in the text) spectra of
resonance states in $^{6}$He.\,\,Energy is reckoned from the ground
state }\label{Fig:SpectrumE&T}
\end{figure}

\noindent In fact, the differential cross section $d\sigma\left(
k,q\right)$ depends only on one parameter, because $k$ and $q$ are
related by condition (\ref{eq:Energy}).\,\,By introducing the energy
of a selected pair of clusters $E_{12}=\frac{\hbar^{2}}{2m}k^{2}$,
we can rewrite the expression for the differential cross section as
\begin{equation}
\frac{d\sigma\left(  E_{12}\right)  }{dE_{12}}\sim\left(
{\displaystyle\int}
\sigma\left(  \mathbf{k},\mathbf{q}\right)  d\widehat{\mathbf{k}}%
d\widehat{\mathbf{q}}\right)  \sqrt{E_{12}\left(  1-E_{12}\right)  }.
\label{eq:CS07}%
\end{equation}

In Figs.\,\,\ref{Fig:CrossSec2PMP} and \ref{Fig:CrossSec1MMP}, we
display the cross sections (\ref{eq:CS07}) for the decay of the
$2^{+}$ and $1^{-}$ resonance states calculated with the MP.\,\,One
can see that the $2^{+}$ resonance state prefers to decay, by
emitting two neutrons with a small value ($E_{12}\approx 0.11$ MeV)
of relative energy and the rather large energy ($E_{12}\approx 0.26$
MeV) of the alpha-particle (with respect to the center of mass of
valence neutrons).\,\,In other  Jacobi coordinates, the resonance
state selects an energy of approximately 0.2~MeV in the nucleon plus
alpha-particle subsystem.\,\,As for the $1^{-}$ resonance state, it
also decays by emitting two neutrons with small relative energy
($E_{12}\approx 0.18$ MeV) and by emitting a neutron and an alpha
particle with the relatively large value of energy ($E_{12}\approx
0.58$ MeV).\,\,The channel $\left( n+\alpha\right) +n$ is dominant
in the decay of the narrow $2^{+}$ resonance state, while the wide
$1^{-}$ resonance state prefers to decay into the
$\left(n+n\right)+\alpha$ channel.\,\,This means, that our model
predicts the sequential decay of the $2^{+}$ resonance state
$^{6}{\rm He}\Rightarrow ^{5}{\rm He}+n\Rightarrow\alpha+n+n$ and
the simultaneous decay of the $1^{-}$ resonance state.

It is worth noting that the results for one-fold cross sections are in
agreement with the results for the correlation functions (Figs.
\ref{Fig:CorrFun1MresonCSMP}, \ref{Fig:CorrFun1MresonMSMP},
\ref{Fig:CorrFun2Preson1CSMP}, and \ref{Fig:CorrFun2Preson1MSMP}),
which show the dominance of the small distance and the small relative momentum
between the valence neutrons in the decay of the $^{6}$He three-cluster
resonance states.

\section{Experiment Versus Theory\label{sect:Compar}}

In this section, we compare the experimental results with the
results of some microscopic and semimicroscopic models.\,\,In Fig.
 \ref{Fig:SpectrumE&T}, we display the theoretical (MP, VP, HH, CSM) and
experimental ($E$) spectra of resonance states in $^{6}$He.\,\,As
the theoretical results, we display the energy of resonance states
calculated within our model with the Volkov potential N2, $m$=0.615,
and the Minnesota potential.\,\,These results are obtained in the
present paper.\,\,We also plotted the spectrum of resonance states
calculated with the hyperspherical harmonics method
\cite{1997PhRvC..55..577D} and with the CSM
\cite{1995PThPh..93...99A}.\,\,We selected those theoretical
calculations, within which not only the well-known $2^{+}$ resonance
state, but other states were discovered.\,\,In Fig.
\ref{Fig:SpectrumE&T}, the symbol $E$ stands for the results of our
experimental investigation with the beam energy $E_{\alpha}=67.2$
MeV.

It is obvious that the first experimental resonance state is the
well-determined $2^{+}$ resonance.\,\,There are two interpretations
of the second experimental resonance state at $E=2.5\pm 0.2$~MeV.
The calculations with the Minnesota potential suggest that this
resonance states is the $J^{\pi}=1^{-}$ state, while our model with
the Volkov potential proposes to interpret it as $J^{\pi}=0^{+}$
state.\,\,Note that the HH and CSM methods have not revealed the
resonance states with the energy close to $E=2.5\pm 0.2$~MeV.\,\,One
sees that the third resonance state ($E=3.1\pm 0.3$~MeV) may be
interpreted as a combination of $2^{+}$ ($E=3.10$~MeV) and $2^{-}$
($E=3.12$~MeV) resonance states, which are obtained with the
MP.\,\,The CSM yields the second $2^{+}$ resonance state with the
energy $E=3.28$ MeV, which is close to the experimental value.
However, there is no resonance states in the HH method with the
energy around $E=3.1\pm 0.3$~MeV.\,\,The resonance states calculated
with the VP lie rather far from the energy $E=3.1\pm 0.3$~MeV.
However, the $2^{+}$ resonance state with $E=3.56$~MeV may be
connected with the third experimental resonance.\,\,Our model
indicates that the wide $1^{-}$ resonance states with the energy
$E=4.26$~MeV (VP) and $E=3.77$~MeV (MP) may be attributed to the
fourth experimental resonance state ($E=4.1\pm 0.3$~MeV).\,\,As for
other theoretical models, the HH model
 predicts two resonance $2^{+}$ and $1^{+}$ states, and the CSM does one $1^{+}$
resonance close to the fourth experimental state.\,\,Concluding this
section, we say that the results of a few theoretical models are in
agreement with those of the present experimental
methods.

\section{Conclusion}

We have considered the resonance structure of $^{6}$He.\,\,Both the
experimental and theoretical methods were used to investigate the
parameters and the nature of resonance states of $^{6}$He.\,\,For
the experimental detection of $^{6}$He resonance states, the
reaction $^{3}$H$(\alpha,p\alpha)2n$ with a four-body exit channel
was used.\,\,The reaction was induced by the interaction of
alpha-particles (the energy of a beam $E_{\alpha} = 67.2$ MeV) with
tritons.\,\,Information about the parameters of resonance states was
obtained from the $p-\alpha$ coincidence spectra.\,\,Ten resonance
states were discovered.\,\,The most part of these states are narrow
resonances, as their total width is less than the energy of a
resonance.

A microscopic model was exploited for the theoretical analysis of
the discrete and continuum spectra in $^{6}$He.\,\,This model
incorporates the dominant three-cluster configuration $\alpha+n+n$
and involves the full set of oscillator functions enumerated by
quantum numbers of the hyperspherical harmonics model to describe
the relative motion of three interacting clusters.\,\,A large set of
hyperradial and hyperspherical excitations was used in calculations
to provide with the stable and convergent results for the energies
and the widths of the resonance states.\,\,Two different effective
$NN$ potentials were used to model the interaction between nucleons
and to determine the interaction between valence neutrons and the
alpha-particle.\,\,We discovered the set of the $0^{+}$, $1^{-}$,
$1^{+}$, and $2^{+}$ resonances.\,\,The total and partial widths of
these resonances were determined, and the dominant decay channels of
the resonances were revealed.\,\,We demonstrated that the resonance
states are mainly formed by one or two channels of the three-cluster
continuum.\,\,These selected channels are weakly coupled to other
channels, which predetermines the existence of resonance states in
the many-channel system.

It was demonstrated that the results of the present microscopic model
satisfactorily agree with the new experimental data.

\vskip-5mm

\rezume{О.М.\,Поворозник, В.С.\,Василевський}{%
СПЕКТР РЕЗОНАНСНИХ СТАНІВ В $^6$He.\\ ЕКСПЕРИМЕНТАЛЬНИЙ ТА
ТЕОРЕТИЧНИЙ АНАЛІЗ} {Вивчено структуру резонансних станів в $^6$He
експериментальними та теоретичними методами. Приведено результати
експериментального дослідження спектра трикластерного континууму
$^6$He.  Для цього залучено реакцію $^3$H$(\alpha,p\alpha)nn$, яка
генерується взаємодією альфа-частинок з тритієм при енергії пучка
$E_{\alpha}=67{,}2$~MеВ. Теоретичний аналіз резонансної структури
$^6$He  проводиться в рамках трикластерної мікроскопічної моделі. Ця
модель використовує гіперсферичні гармоніки для опису динаміки
відносного руху кластерів. Набір нових резонансних станів знайдено
експериментальним та теоретичним методами. Визначено енергію, ширину
та домінуючі канали розпаду резонансів. Отримані результати детально
порівнюються із результатами різних теоретичних моделей, а також
експериментів. }

\end{document}